\documentclass[twoside,twocolumn,9pt]{article}
\usepackage{extsizes}
\usepackage[super,sort&compress,comma]{natbib} 
\usepackage[version=3]{mhchem}
\usepackage[left=1.5cm, right=1.5cm, top=1.785cm, bottom=2.0cm]{geometry}
\usepackage{balance}
\usepackage{times,mathptmx}
\usepackage{sectsty}
\usepackage{graphicx} 
\usepackage{lastpage}
\usepackage[format=plain,justification=justified,singlelinecheck=false,font={stretch=1.125,small,sf},labelfont=bf,labelsep=space]{caption}
\usepackage{float}
\usepackage{fancyhdr}
\usepackage{fnpos}
\usepackage[english]{babel}
\addto{\captionsenglish}{%
  
}
\usepackage{array}
\usepackage{droidsans}
\usepackage{charter}
\usepackage[T1]{fontenc}
\usepackage[usenames,dvipsnames]{xcolor}
\usepackage{setspace}
\usepackage[compact]{titlesec}
\usepackage{hyperref}
\usepackage{amsmath}
\usepackage{amsfonts}
\usepackage{amssymb}
\usepackage{bm}

\newcommand*{\citen}[1]{%
  \begingroup
    \romannumeral-`\x % remove space at the beginning of \setcitestyle
    \setcitestyle{numbers}%
    \cite{#1}%
  \endgroup
}

\DeclareMathAlphabet\mathzapf       {T1}{pzc} {mb} {it}

\usepackage[normalem]{ulem}
\usepackage{scalerel}

\definecolor{cream}{RGB}{222,217,201}

\begin{document}

\pagestyle{plain}
\thispagestyle{plain}

%%%PAGE SETUP - Please do not change any commands within this section%%%
\makeFNbottom
\makeatletter
\renewcommand\LARGE{\@setfontsize\LARGE{15pt}{17}}
\renewcommand\Large{\@setfontsize\Large{12pt}{14}}
\renewcommand\large{\@setfontsize\large{10pt}{12}}
\renewcommand\footnotesize{\@setfontsize\footnotesize{7pt}{10}}
\makeatother

\renewcommand{\thefootnote}{\fnsymbol{footnote}}
\renewcommand\footnoterule{\vspace*{1pt}% 
\color{cream}\hrule width 3.5in height 0.4pt \color{black}\vspace*{5pt}} 
\setcounter{secnumdepth}{5}

\makeatletter 
\renewcommand\@biblabel[1]{#1}
\renewcommand\@makefntext[1]% 
{\noindent\makebox[0pt][r]{\@thefnmark\,}#1}
\makeatother 
\renewcommand{\figurename}{\small{Fig.}~}
\sectionfont{\sffamily\Large}
\subsectionfont{\normalsize}
\subsubsectionfont{\bf}
\setstretch{1.125} %In particular, please do not alter this line.
\setlength{\skip\footins}{0.8cm}
\setlength{\footnotesep}{0.25cm}
\setlength{\jot}{10pt}
\titlespacing*{\section}{0pt}{4pt}{4pt}
\titlespacing*{\subsection}{0pt}{15pt}{1pt}
%%%END OF PAGE SETUP%%%

%%%FOOTER%%%
\fancyfoot{}
\fancyfoot[RO]{\footnotesize{\sffamily{1--\pageref{LastPage} ~\textbar  \hspace{2pt}\thepage}}}
\fancyfoot[LE]{\footnotesize{\sffamily{\thepage~\textbar\hspace{3.45cm} 1--\pageref{LastPage}}}}
\fancyhead{}
\renewcommand{\headrulewidth}{0pt} 
\renewcommand{\footrulewidth}{0pt}
\setlength{\arrayrulewidth}{1pt}
\setlength{\columnsep}{6.5mm}
\setlength\bibsep{1pt}
%%%END OF FOOTER%%%

%%%FIGURE SETUP - please do not change any commands within this section%%%
\makeatletter 
\newlength{\figrulesep} 
\setlength{\figrulesep}{0.5\textfloatsep} 

\newcommand{\topfigrule}{\vspace*{-1pt}% 
\noindent{\color{cream}\rule[-\figrulesep]{\columnwidth}{1.5pt}} }

\newcommand{\botfigrule}{\vspace*{-2pt}% 
\noindent{\color{cream}\rule[\figrulesep]{\columnwidth}{1.5pt}} }

\newcommand{\dblfigrule}{\vspace*{-1pt}% 
\noindent{\color{cream}\rule[-\figrulesep]{\textwidth}{1.5pt}} }

\makeatother
%%%END OF FIGURE SETUP%%%

%%%TITLE, AUTHORS AND ABSTRACT%%%
\twocolumn[
  \begin{@twocolumnfalse}
  {\fontsize{14.5}{0} \selectfont \textbf{Emergent vortices and phase separation in systems of chiral active particles with dipolar interactions}} \\
  \\
  \large{Guo-Jun Liao$^{\ast}$\textit{$^{a}$} and Sabine H. L. Klapp$^{\ast}$\textit{$^{a}$}} \\
  \\
  \normalsize{%
Using Brownian dynamics (BD) simulations we investigate the self-organization of a monolayer of chiral active particles with dipolar interactions. 
Each particle is driven by both, translational and rotational self-propulsion, and carries a permanent point dipole moment at its center. 
The direction of the translational propulsion for each particle is chosen to be parallel to its dipole moment.
Simulations are performed at high dipolar coupling strength and a density below that related to motility-induced phase separation in simple active Brownian particles.
Despite this restriction, we observe a wealth of phenomena including formation of two types of vortices, phase separation, and flocking transitions.
To understand the appearance and disappearance of vortices in the many-particle system, we further investigate the dynamics of simple ring structures under the impact of self-propulsion.
} \\
 \end{@twocolumnfalse} \vspace{0.6cm}

  ]
%%%END OF TITLE, AUTHORS AND ABSTRACT%%%

%%%FONT SETUP - please do not change any commands within this section
\renewcommand*\rmdefault{bch}\normalfont\upshape
\rmfamily
\section*{}
\vspace{-1cm}

%%%FOOTNOTES%%%

\footnotetext{\textit{$^{a}$~
Institut f\"ur Theoretische Physik, Technische Universit\"at Berlin, Hardenbergstr. 36, D-10623 Berlin, Germany. E-mail: guo-jun.liao@campus.tu-berlin.de, klapp@physik.tu-berlin.de}}

%Please use \dag to cite the ESI in the main text of the article.

%%%END OF FOOTNOTES%%%

%%%MAIN TEXT%%%%
\section{\label{DCSs:sec:intro}Introduction}
Systems of self-propelled particles consist of a large number of motile constituents, each of which is capable of continuously converting energy from an internal source or the surroundings into mechanical motion.\cite{Romanczuk2012, Zottl2016}
Examples of biological self-propelled particles can be found over a wide range of length and time scales, from bird flocks, fish schools, mammalian herds, and pedestrian crowds in our daily life, to bacteria, sperm cells, and microtubules at the microscale.\cite{Ramaswamy2010, Bechinger2016}
Self-propelled particles can also be synthesized in the laboratory, famous examples including bimetallic nanorods,\cite{Paxton2004, Paxton2006} spherical Janus particles,\cite{Gangwal2008, Buttinoni2012} and magnetic rollers.\cite{Kaiser2017, Kokot2018, Han2020}
It is now well established that already relatively simple systems of self-propelled particles can display complex collective behavior, giving rise to a still increasing scientific interest.\cite{Gompper2020}
A prominent example of such complex behavior is motility-induced phase separation (MIPS) which occurs in systems of disk-shaped, active Brownian particles above a critical density $\Phi_{crit}$.\cite{Tailleur2008, Buttinoni2013, Cates2015, Speck2015, Digregorio2018}
Another ``classical'' example is the flocking transition in the Vicsek model,\cite{Vicsek1995} a system of self-propelled point-like particles with ferromagnetic interactions.
%

%% New Paragraph
%
%% Introduction of circle swimmers
The majority of theoretical and numerical studies on active particles assumes that each particle tends to ``swim'' along a straight line due to the translational self-propulsion.
%
%However, this assumption clearly becomes invalid once a rotational self-propulsion comes into play.
However, this assumption clearly becomes invalid once the chiral symmetry along the propelling direction is broken.
Such broken symmetry often introduces a rotational self-propulsion, which, together with a translational self-propulsion, causes a single swimmer to move along a perfect circular path in the absence of thermal noise.\cite{VanTeeffelen2008}
Hence, active particles simultaneously subject to the rotational and translational self-propulsion are generally referred to as ``chiral active particles,''\cite{Bechinger2016} or  ``circle swimmers.''\cite{Lowen2016}
Famous examples include \textit{E. coli} cells close to a surface,\cite{DiLuzio2005, Lauga2006} FtsZ proteins,\cite{Loose2014, Denk2016, Ramirez-Diaz2018} and synthetic L-shaped particle.\cite{Kummel2013}
Recent research has shown that chiral motion (\textit{i.e.}, circle swimming) can indeed significantly change the self-assembly dynamics in active systems.
Already for the simplest models of chiral active particles, it has been shown analytically and numerically that circle swimming generally suppresses MIPS,\cite{Bickmann2020} induces clockwise vortices,\cite{Liao2018} and yields hyperuniform structures with vanishing long-wavelength (but large local) density fluctuations.\cite{Lei2019}
Recent studies of circle swimmers with short-range anisotropic interactions have revealed even more intriguing features.
For example, simulations have demonstrated that chiral active particles with an asymmetric\cite{Yang2014, Denk2016, Zhang2020} or elongated\cite{Kaiser2013} shape display vortex structures. 
Chiral active particles with chemotactic alignment can form a global traveling wave.\cite{Liebchen2016}
Further, chiral active particles with polar (\textit{i.e.}, ferromagnetic Heisenberg-like) interactions exhibit rotating macrodroplets,\mbox{\cite{Liebchen2017, Levis2018}} chimera states,\mbox{\cite{Kruk2018, Kruk2020, Kruk2020a}} chiral self-recognition,\cite{Arora2021} dynamical frustration,\cite{Huang2020} or a surface-population reversal effect in ring-shaped confinement.\cite{Fazli2021}
Additional effects can arise in the presence of a rotating magnetic field, where disk-shaped chiral active particles without anisotropic interactions can display chiral separation and polar ordering.\cite{Lin2020}
In the present study, we go one step further and consider chiral active particles with dipole-dipole interactions.
These interactions differ from the previously considered anisotropic ones by the longer range and by a more complicated angular dependency.
That is, the pair interaction depends not only on the angle between the two involved dipole moments, but also on their spatial configuration which eventually promotes chain formation. 
To explore the role of these factors for the resulting collective behavior is interesting not only from an academic point of view. 
Rather, many microswimmers have embedded (permanent or induced) magnetic or electric dipole moments giving rise to dipole-dipole interactions, examples being metallodielectric Janus spheres,\cite{Gangwal2008, Kogler2015} magnetotactic bacteria,\cite{Meng2018, Klumpp2019} and magnetic rollers.\cite{Kaiser2017, Kokot2018, Han2020}
Therefore, the overall aim of this study is to enhance the understanding of the active matter systems involving circle swimming and dipole-dipole interactions.
To this end, we consider a two-dimensional system of chiral active particles with dipolar interactions, each of which moves at a self-propulsion speed (characterized by the particle motility) and rotates at an ``active'' angular speed.
The (permanent) dipole moment embedded in each particle is assumed to be directed along the direction of translational self-propulsion.
Our model combines features of non-dipolar, disk-shaped circle swimmers\cite{Liao2018} and dipolar active Brownian particles.\cite{Liao2020}
Therefore, we refer to our model as dipolar circle swimmers in the rest of this paper.
Based on Brownian dynamics simulations, we explore a wide range of the particle motility and the active angular speed at a large dipolar coupling strength and a low density.
%

%% New paragraph
%
As a unique feature of dipolar circle swimmers, we observe two types of vortices, which we refer to as Type I and Type II vortices. 
Type I vortices consist of forward-moving particles whose dipole moments display counterclockwise patterns, while Type II vortices are backward-moving particles whose dipole moments show clockwise structures.
As the motility increases from zero, Type II vortices vanish, and the system exhibits significant phase separation with the dense domain characterized by giant Type I vortices.
We show that some aspects of the vortex formation can be explained by considering the dynamics of simple ring structures.
At even higher motilities, the system displays flocking patterns, in which dipolar circle swimmers self-assemble into polar clusters and show a significant global orientational order.
The rest of this paper is organized as follows. 
In Sec.~\ref{DCSs:sec:method} we describe our methods of investigation, including the model, the simulation setup, and the target quantities.
An overview of the system behavior (at a fixed density) is given in Sec.~\ref{DCSs:sec:result4}. 
In Sec.~\ref{DCSs:sec:result1}~$-$~\ref{DCSs:sec:result3}, we discuss specific aspects such as chain formation, clustering and phase separation, emergence of vortices, and orientational ordering.
Finally, we summarize our findings and provide an outlook in Sec.~\ref{DCSs:sec:conclusions}.

\section{\label{DCSs:sec:method}Model and methods of investigation}
Our model system of dipolar circle swimmers combines the previously investigated model of the dipolar active Brownian particles\cite{Liao2020} and the model of spherical circle swimmers.\cite{VanTeeffelen2008, Mijalkov2013, Ao2015, Jahanshahi2017, Liao2018, Lei2019}
Therefore, various methods of investigation can be transferred from previous works.
In the following, we summarize the main points and refer the reader for details to ref.~\citen{Liao2018, Liao2020}.

\subsection{Model}
We consider $N$ disk-shaped Brownian particles with diameter $\sigma$ dispersed in a monolayer in the $xy$-plane.
Each of the particles carries a permanent point dipole moment $\boldsymbol{\mu}_i$ located at its center.
In addition, each particle is subject to a self-propulsion force $F_0 \hat{\boldsymbol{e}}_i$ ($i = 1, \ldots, N$) and torque $M_0 \hat{\boldsymbol{z}}$.
The particle orientation $\hat{\boldsymbol{e}}_i$ denotes the direction of self-propulsion force and is assumed to be directed along the unit dipole moment $\hat{\boldsymbol{\mu}}_i $ at each instant of time, \textit{i.e.}, $\hat{\boldsymbol{e}}_i = \hat{\boldsymbol{\mu}}_i$.
The pairwise interaction potential for two swimmers $i$ and $j$ is given by
\begin{equation} \label{DCSs:eqn:pair}
  u_{pair}\big(\boldsymbol{r}_{ij}, 
                \boldsymbol{\mu}_i, 
                \boldsymbol{\mu}_j \big) = 
    u_{sr}\left(r_{ij}\right) + 
    u_{dd}\big(\boldsymbol{r}_{ij}, 
                \boldsymbol{\mu}_i, 
                \boldsymbol{\mu}_j \big)
      \text{,}
\end{equation}
where $u_{sr}\left(r_{ij}\right)$ with $r_{ij} = \vert \boldsymbol{r}_{ij} \vert = \vert \boldsymbol{r}_{j} - \boldsymbol{r}_{i} \vert$ stands for the short-range (sr) steric repulsion, which prevents particles from overlapping.
Specifically, we employ the Weeks-Chandler-Anderson potential\cite{Weeks1971} defined as
\begin{equation} \label{DCSs:eqn:wca}
u_{sr}(r_{ij})= 
  \begin{cases}
    4\epsilon \left[
      \left(\dfrac{\sigma}{r_{ij}}\right)^{12} - 
      \left(\dfrac{\sigma}{r_{ij}}\right)^{6} + 
      \dfrac{1}{4}
    \right]
    \text{,} &\text{if $r_{ij} < r_{c}$,}\\
    0\text{,} &\text{else.}
  \end{cases}
\end{equation}
The potential is truncated at a cut-off (c) distance $r_{c} = 2^{1/6}\sigma$, with $\sigma$ being the particle diameter.
The repulsive strength is described by $\epsilon^* = \beta \epsilon$ with $\beta^{-1} = k_B T$ representing the thermal energy (with $k_B$ denoting the Boltzmann constant and $T$ the temperature).
The second term on the right-hand side of Eq.~\eqref{DCSs:eqn:pair} represents the (long-range) dipole-dipole (dd) interaction, given by
\begin{equation} \label{DCSs:eqn:Udd}
u_{dd}\big(\boldsymbol{r}_{ij}, \boldsymbol{\mu}_{i}, \boldsymbol{\mu}_{j}\big)= 
      \dfrac{\boldsymbol{\mu}_{i} \cdot \boldsymbol{\mu}_{j}}{r_{ij}^{3}} - 
      3 \dfrac{
          \big(\boldsymbol{\mu}_{i} \cdot \boldsymbol{r}_{ij} \big)
          \big(\boldsymbol{\mu}_{j} \cdot \boldsymbol{r}_{ij} \big)
        }
      {r_{ij}^{5}}
    \text{.}
\end{equation}
The strength of the dipole-dipole interaction is characterized by the parameter $\lambda = \beta\mu^2\sigma^{-3}$, where $\mu = \vert \boldsymbol{\mu}_{i} \vert$ denotes the strength of each dipole moment.

\subsection{\label{DCSs:sec:EoM}Equations of motion}
To investigate the dynamical self-assembly of dipolar circle swimmers, we perform extensive Brownian dynamics (BD) simulations involving $N$ particles in a squared box with side length $L$.
The motion of the $i$th particle is described by the coupled Langevin equations in the overdamped limit,\cite{VanTeeffelen2008}
\begin{align}
  \dot{\boldsymbol{r}}_i & = 
    \beta D_t
      \Big[
        F_{0}\widehat{\boldsymbol{e}}_{i} 
        - \nabla_{\boldsymbol{r}_i}U_{i} 
        + \boldsymbol{\xi}_{i}\left(t\right) 
      \Big]\text{,} \label{DCSs:eqn:coupled_Langevin_trans}\\
  \dot{\psi}_i & = 
    \beta D_r
      \Big[
        M_{0}
        - \partial_{\psi_i} U_{i} 
        + \Gamma_{i}\left(t\right)
      \Big]\text{,} \label{DCSs:eqn:coupled_Langevin_rot}
\end{align}
where $\boldsymbol{r}_{i}$ denotes the particle position and $\psi_{i}$ represents the polar angle for the orientation $\hat{\boldsymbol{e}}_i = \left( \text{cos}\psi_{i}, \text{sin}\psi_{i} \right)$.
Focusing on $\psi_{i}$ as the relevant angle, we implicitly assume that the dipole moment $\boldsymbol{\mu}_{i}$ are oriented along in-plane directions (for justification, see Sec.~\ref{DCSs:sec:parameter}).
The dots above $\boldsymbol{r}_{i}$ and $\psi_{i}$ on the left-hand side of Eq.~\eqref{DCSs:eqn:coupled_Langevin_trans} and~\eqref{DCSs:eqn:coupled_Langevin_rot} indicate time derivatives, and the potential energy for the $i$th particle $U_{i}$ is the sum over all the pairwise potentials between particle $i$ and all the other particles $j$, \textit{i.e.},
\begin{equation} \label{DCSs:eqn:ui}
  U_{i} = \sum_{j = 1, j \neq i}^{N} 
            u_{pair}\big(\boldsymbol{r}_{ij},
                         \boldsymbol{\mu}_{i},
                         \boldsymbol{\mu}_{j}
                    \big)\text{,}
\end{equation}
with $u_{pair}$ being the pair potential [see Eq. \eqref{DCSs:eqn:pair}].
In Eq.~\eqref{DCSs:eqn:coupled_Langevin_trans}, $D_t$ denotes the translational diffusion constant (we do not use the tensorial quantity due to the disk-like shape of the particles).
Correspondingly, for the rotational motion, $D_r$ describes the rotational diffusion constant.
To account for the Brownian motion, the random force $\boldsymbol{\xi}_{i}\left(t\right)$ and torque $\Gamma_{i}\left(t\right)$ are Gaussian white noises, which have zero means and are delta correlated,
$\langle
   \boldsymbol{\xi}_{i}(t)
\rangle = \boldsymbol{0}$,
$\langle
  \Gamma_i(t)
\rangle = 0$,
$\langle
   \boldsymbol{\xi}_{i}(t)
   \otimes
   \boldsymbol{\xi}_{j}(t')
 \rangle =
   2  \delta_{ij}  
      \delta(t-t')
      \mathbb{I} /
        (D_t\beta^{2})$, 
and 
$\langle
  \Gamma_i(t)
  \Gamma_j(t')
\rangle = 
  2  \delta_{ij}
     \delta(t-t') /
       (D_r\beta^{2})$.
Here, the angle brackets $\langle \cdots \rangle$ stand for ensemble average, and $\otimes$ denotes dyadic product.
Finally, it is worth recalling the behavior of a single circle swimmer (without interacting with other swimmers).
In the absence of thermal fluctuations and at $M_0 > 0$, the swimmer moves along a closed circular path of radius $R = D_t F_0 / \left( D_r M_0\right)$ and rotates \textit{counterclockwise} at a self-propulsion speed $v_0= \beta D_t F_0$ and angular speed $\omega_0 = \beta D_r M_0$.\cite{VanTeeffelen2008}

\subsection{\label{DCSs:sec:parameter}Parameter choice}
Following ref.~\citen{Liao2020}, we choose the strength of steric repulsion as $\epsilon^* = 10$ and $D_r = 2.57914 D_t / \sigma^2$.
The dipolar coupling strength is set to $\lambda = 10$, such that the dipolar pair energy of two particles at contact is ten times the thermal energy.
To model chain formation and orientational ordering in dipolar colloids, one typically considers dipolar coupling strengths between $\lambda \approx 1$ and $\lambda \approx 10$.\mbox{\cite{Weis2005, Schmidle2011, Jager2012}} 
However, test simulations of the present model with $\lambda$ varying from $0$ to $10$ show that vortices (which will be discussed in Sec.~\ref{DCSs:sec:result5}) only appear once $\lambda \approx 10$. 
A further increase of $\lambda > 10$ requires even smaller time steps $\delta t$, which, together with the long-range feature of dipole-dipole interactions, makes the simulations even more computationally expensive. 
Therefore, in the present work the dipolar coupling strength is fixed to $\lambda = 10$.
For passive monolayers of such strongly coupled dipolar particles, it is well known that there is a pronounced tendency to orient along in-plane directions; specifically, one observes self-assembly into chains and rings,\cite{Tavares2002, Duncan2004, Duncan2006, Kantorovich2008, Cerda2008} or dense planar ordered states.\cite{Tavares2002, Klapp2002, Ouyang2011, Geiger2013}
Therefore, we here assume beforehand that the dipole moment $\boldsymbol{\mu}_i$ lies in the $xy$-plane, \textit{i.e.}, the $z$-component is neglected.
Based on this assumption, a two-dimensional Ewald summation is employed to deal with the long-range feature of dipole-dipole interactions, as outlined in the Appendix of ref.~\citen{Liao2020}.
Equations~\eqref{DCSs:eqn:coupled_Langevin_trans} and~\eqref{DCSs:eqn:coupled_Langevin_rot} are solved numerically \textit{via} the Euler$-$Maruyama method\cite{Kloeden1992} with a discrete time step $\delta t = 2 \times 10^{-5} \tau$, where $\tau = \sigma^2/D_t$ denotes the Brownian diffusive time.
We define the particle diameter $\sigma$, the thermal energy $k_B T$, and the Brownian diffusive time $\tau$ as the units of length, energy, and time, respectively.
All physical quantities in the system are then expressed in the units based on the
dimensional combination of these three basic units.
Consistent with our earlier work,\cite{Liao2018} the impact of the translational and rotational self-propulsion with respect to the thermal noise is represented by the dimensionless motility $v_0^* = v_0 \sigma/ D_t$ and the active angular speed $\omega_0^* = \omega_0 / D_r$.
Moreover, we set the particle number to $N = 1156$, unless otherwise stated.
The sizes of the present simulations are limited to the order of $10^3$ particles, due to the expensive computational cost resulting from the long-range feature of dipole-dipole interactions and the necessity of using small time step $\delta t = 2 \times 10^{-5} \tau$ for simulating active particles with strong steric repulsion ($\epsilon^* = 10$).
As the initial configuration for all simulations, we distribute particles on a square lattice and assign a random orientation to each of them.
For reaching a steady state, we typically wait at least $5 \times 10^5$ steps.
Then, we start to measure the statistical properties (see Sec.~\ref{DCSs:sec:target_quant}) every $500$ steps and collect at least $1000$ samples for each realization.
Our main focus of this study is on the pattern formation of dipolar circle swimmers in the regime of low densities.
Specifically, the mean area fraction is chosen to be $\Phi = N \pi \sigma_{eff}^2 / \left(4 L^2\right) = 0.23$, where the effective hard disk diameter is given by $\sigma_{eff} \approx 1.07851 \sigma$ (see ref.~\citen{Liao2020} for details).
We note that the density considered here is below the critical density of motility-induced phase separation (MIPS), $\Phi_{crit}$, in purely repulsive systems, which is in the range $0.28 \lesssim \Phi_{crit} \lesssim 0.4$.\cite{Bruss2018, Maloney2020}
In other words, at $\Phi = 0.23$ and in the absence of dipole-dipole interactions, MIPS does not occur even at extremely high motilities $v_0^*$.

\subsection{\label{DCSs:sec:target_quant}Observables}

In this section we discuss the relevant observables and their numerical analysis (more details can be found in ref.~\citen{Liao2018, Liao2020}).

\begin{table*}[!bhtp]
  \centering
  \small
  \begin{tabular}{ l || c | c | c } 
        & Chain & Clustering, phase separation & Orientational \\
  State & formation & and emergent vortices & ordering\\
  \hline\hline
  Percolated networks & $\Pi > 0.5$, $\phi_p > 0.5$ & $d_e < 3$ & $\phi_{\boldsymbol{e}} \leq 0.5$ \\
  Chain-like structures & $\Pi \leq 0.5$, $\phi_p > 0.5$ & $d_e < 3$ & $\phi_{\boldsymbol{e}} \leq 0.5$\\ 
  Finite-size vortices & - & $d_e \geq 3$, single peak in $P\left(\phi\right)$ & $\phi_{\boldsymbol{e}} \leq 0.5$\\
  Vortices with phase separation & - & $d_e \geq 3$, double peaks in $P\left(\phi\right)$ & $\phi_{\boldsymbol{e}} \leq 0.5$\\
  Micro-flocking & - & monotonic decay of $nP\left(n\right)$ & $\phi_{\boldsymbol{e}} > 0.5$\\
  Macro-flocking & - & broad shoulder or a peak  & $\phi_{\boldsymbol{e}} > 0.5$\\
  & & in $nP\left(n\right)$ at large $n$ & \\
  \end{tabular}
  \caption{Characterization of the states of dipolar circle swimmers according to the target quantities described in Sec.~\ref{DCSs:sec:target_quant}.}
  \label{DCSs:table:states}
\end{table*}

\subsubsection{\label{DCSs:sec:chaining}Percolation and polymerization}

It is well established that passive dipolar spheres with strong dipole-dipole coupling have a tendency to aggregate into chains, rings, and percolated networks.\cite{Laria1991, Weis1993, Duncan2004, Weis2005, Duncan2006, Rovigatti2011, Rovigatti2012, Rovigatti2013, Kantorovich2013, Kantorovich2016, Ronti2017, Camp2018}
Therefore, we expect similar patterns to emerge in the present system, at least at low motilities and angular speeds.
A first measure is the percolation probability $\Pi$, defined as the probability of finding a cluster in a simulation snapshot that connects two opposite sides of the simulation box.\cite{Laria1991}
Following earlier studies of active colloidal systems,\mbox{\cite{Buttinoni2013, Sese-Sansa2018, VanDerLinden2019}} a cluster is defined as follows: 
Two particles are regarded as being associated if their center-to-center distance is smaller than a ``critical'' distance $r_L$.
A cluster is then a set of particles that are mutually associated.
Specifically, we set $r_L = 1.2\sigma$.
As it turns out, the results do not change significantly if we choose a different value of $r_L$, as long as it corresponds to a distance between the location of the first peak and the first valley of the pair correlation function for the corresponding ``reference system'' ($v_0^* = 0$, $\omega_0^* = 0$, and $\lambda = 0$) at $\Phi = 0.23$.
Further, for the present system of dipolar circle swimmers, the location of the first peak and the first valley do not vary too much upon variation of $v_0^*$ and $\omega_0^*$. 
Therefore, we fix $r_L = 1.2\sigma$ throughout this work for simplicity.

The percolated dipolar networks may break, \textit{e.g.}, upon an increase of temperature.\cite{Rovigatti2012}
In this case, the system may display a ``string'' fluid state with many ``polymerized'' chains, which are composed of dipolar spheres connecting with their neighbors in a head-to-tail fashion.
In self-assembly studies of dipolar systems, these chain structures are commonly quantified by the degree of ``polymerization,''\cite{Rovigatti2012, Liao2020, Maloney2020, Maloney2020a}
\begin{equation} \label{DCSs:eqn:phi_p}
  \phi_p = \left\langle N_p \right\rangle / N \text{.}
\end{equation}
In the above equation, $N_p$ denotes the number of particles in the ``polymerized'' chains.
In the present study, a chain is defined as a cluster comprised of at least ten particles, which conforms with the large value of the dipolar coupling strength ($\lambda = 10$).
We note that the above definition of a chain cannot distinguish between an elongated chain and a compact disk-shaped cluster with local head-to-tail ordering.
This problem cannot be solved by adding more complex criteria to the definition of clusters (as ref.~\citen{Liao2020} does), \textit{i.e.}, $\boldsymbol{\mu}_i \cdot \boldsymbol{\mu}_j > 0$ and $(\boldsymbol{\mu}_i \cdot \boldsymbol{r}_{ij})(\boldsymbol{\mu}_j \cdot \boldsymbol{r}_{ij}) > 0$, since head-to-tail ordering is present in all types of aggregates. 
The degree of polymerization $\phi_p$ defined in Eq.~\eqref{DCSs:eqn:phi_p} approaches unity if all particles self-assemble into string-like chains or disk-shaped clusters, and is zero if there is no chain structure.
We consider the system to display a state with polymerized string fluids if $\phi_p > 0.5$, and no giant disk-shaped clusters (to be discussed later) are present.

\subsubsection{Clustering and phase separation}
As mentioned in Sec.~\ref{DCSs:sec:parameter}, the density considered in this work is below the critical density of the MIPS occurring in non-dipolar active Brownian particles.
Nevertheless, it has been shown that some anisotropic (\textit{e.g.}, polar) interactions can enhance the tendency for MIPS.\cite{Pu2017, Sese-Sansa2018}
It thus seems worth checking for this phenomenon also in the present system.
As a first step, we investigate the clustering behavior by measuring the fraction of the largest cluster\cite{Buttinoni2013, Speck2014, Speck2015, Liao2018, Sese-Sansa2018, Liao2020}
\begin{equation}
  \label{DCSs:eqn:phi_c}
  \phi_c = \dfrac{\left\langle n_{lc} \right\rangle}{N}\text{,}
\end{equation}
where $n_{lc}$ denotes the size of the largest cluster.
The definition of a cluster is given in Sec.~\ref{DCSs:sec:chaining}.
The order parameter $\phi_c$ reaches unity if the largest cluster is composed of all swimmers, and approaches zero if swimmers are homogeneously distributed.
For the present dipolar system, we note that both, the fraction of the largest cluster $\phi_c$ and the degree of polymerization $\phi_p$ (see Eq.~\eqref{DCSs:eqn:phi_p}), are close to unity if a single giant cluster is present.
On the other hand, $\phi_c \ll 1$ and $\phi_p \approx 1$ indicate that the system forms many chains with intermediate size.
We also note that a large value of $\phi_c$ alone cannot distinguish between a state with percolated networks and a state with giant disk-shaped clusters. 
To solve this, we monitor, in addition, the percolation probability $\Pi$.
Percolated networks are identified by $\phi_c > 0.5$ and $\Pi > 0.5$, while giant compact clusters are identified by $\phi_c > 0.5$ and $\Pi \leq 0.5$.
Finally, to identify phase separation, we employ a Voronoi tessellation to obtain the probability distribution function $P\left(\phi\right)$ of the local area fraction $\phi$ without a short-time average (see ref.~\citen{Liao2020} for details). 
The system is regarded to display phase separation if the density profile $P\left(\phi\right)$ shows a double-peak structure. 
Subsequently, the coexisting densities are the density values corresponding to the peaks.
\subsubsection{\label{DCSs:sec:vortex}Emergent vortices}

For active systems on a surface, a vortex structure is generally defined as a disk-shaped high-density region in which the particle velocities (or the coarse-grained velocity field) display circular patterns with a common center.
Vortices emerge, for instance, in various systems composed of biological motile constituents, including microtubules\cite{Sumino2012} and bacterial suspensions.\cite{Wensink2012, Wioland2013, Reinken2020}
Further, biological circle swimmers, such as FtsZ filaments,\cite{Loose2014, RamirezDiaz2017} and spermatozoa of sea urchins,\cite{Riedel2005} are found to be able to self-organize into an array of vortices.
Vortex patterns can also be observed in synthetic colloidal systems of magnetic rollers.\cite{Kaiser2017, Kokot2018, Han2020}
Given these examples, it seems possible that also the present system of dipolar circle swimmers may self-organize into vortices, owing to the interplay between the active rotation and the dipole-dipole interactions.
Inspired by ref.~\citen{Wensink2012}, we characterize the vortex structures by analyzing the orientational correlation function (recall that the direction of the particle's dipole moment coincides with its orientation, $\hat{\boldsymbol{\mu}}_i = \hat{\boldsymbol{e}}_i$),
\begin{equation} \label{DCSs:eqn:Ce}
  C_{\boldsymbol{e}}\left(r\right) = 
    \dfrac{
      \left\langle
        \sum_{ i, j}
        \hat{\boldsymbol{\mu}}_i \cdot
        \hat{\boldsymbol{\mu}}_j
        \delta \left( r - r_{ij} \right)
      \right\rangle
    }
    {
      \left\langle
        \sum_{ i, j}
        \delta \left( r - r_{ij} \right)
      \right\rangle
    }\text{.}
\end{equation}
If $i = j$, $r_{ij} = \vert \boldsymbol{r}_{j} - \boldsymbol{r}_{i} \vert =  \vert \boldsymbol{r}_{i} - \boldsymbol{r}_{i} \vert = 0$ and $\hat{\boldsymbol{\mu}}_i \cdot \hat{\boldsymbol{\mu}}_j = \hat{\boldsymbol{\mu}}_i \cdot \hat{\boldsymbol{\mu}}_i = 1$. 
Since particles do not overlap, we have $r_{ij} > 0$ for $i \neq j$. 
Thus, we obtain $C_{\boldsymbol{e}}\left(r = 0\right) = 1$.
If particle $i$ and $j$ are infinitely far apart ($r_{ij} \to +\infty$), we assume that their average orientational correlation is very weak and is thus approximately zero at the given low density.
Since $C_{\boldsymbol{e}}\left(0\right) = 1$ and $\lim_{r \to +\infty} C_{\boldsymbol{e}}\left(r\right) \approx 0$, a negative correlation at a finite distance $r$ represents that particle orientations (\textit{i.e.}, propulsion directions) on average point in the opposite directions when separated by the distance $r$.
Hence, the system is regarded to self-assemble into vortex structures if $C_{\boldsymbol{e}}\left(r\right)$ shows negative values at finite distances.
The distance corresponding to the minimum of $C_{\boldsymbol{e}}(r)$ can be used to characterize the average diameter of vortices $d_{\boldsymbol{e}}$, defined by 
\begin{equation}
  \label{DCSs:eqn:de}
  \dfrac{dC_{\boldsymbol{e}}(r)}{dr} \Bigg\vert_{\scaleto{r = d_{\boldsymbol{e}}}{9pt} } = 0\text{.}
\end{equation}
%
%Further, we take the minimum value of the correlation function $C_{\boldsymbol{e}}(d_{\boldsymbol{e}})$ to describe the prominence of the vortex structures.
%

\subsubsection{Flocking behavior}

Orientational ordering is commonly observed in a variety of active systems with anisotropic interactions.
As a prominent example, the Vicsek model describes self-propelled particles whose velocities tend to align with those of their neighbors when perturbed by noise.\cite{Vicsek1995}
At high motilities, this model typically exhibits the so-called flocking behavior, in which particles self-organize into clusters with significant polar ordering and move collectively toward a certain direction. 
Moreover, the flocking states are found to persist in the presence of active rotation\cite{Liebchen2017, Levis2018} and steric repulsion.\cite{Martin-Gomez2018}
Although the dipole-dipole interactions in the present model are more complex than the Heisenberg-like interactions in the Vicsek model, we still expect to observe flocking patterns in the present system.
This expectation is also motivated by our observation of flocking in dipolar active Brownian particles.\cite{Liao2020}
We characterize the emergence of flocking behavior by the global polar order parameter,\cite{Ginelli2016}
\begin{equation}
  \phi_{\boldsymbol{e}}
  = 
  \dfrac{1}{N} 
  \left\langle 
    \left\vert 
      \sum_{i=1}^{N}
      \hat{\boldsymbol{e}}_i
    \right\vert
  \right\rangle \text{.} \label{DCSs:eqn:phi_e}
\end{equation}
This order parameter represents the magnitude of the average orientation, which reaches unity if all particles move toward the same direction, and zero if the particle orientations are not correlated.
Depending on the length scale of polar clusters with respect to the particle
number $N$, the flocking states can be further classified into micro- and macro-flocking
states.
To this end, we follow the treatment in ref.~\citen{Liao2020} and measure the distribution function of the cluster size $P\left(n\right)$, which is the probability that a randomly selected cluster is composed of $n$ particles.
In the case of macro-flocking, the system is usually composed of several large polar clusters with many small polar clusters, such that $P\left(n\right)$ is extremely small at large $n$.
To solve this problem, we consider the weighted distribution function $nP\left(n\right)$, whose value is proportional to the probability that a randomly selected particle belongs to a chain with a size $n$.
We identify the swimmers are in a micro-flocking state if $\phi_{\boldsymbol{e}} > 0.5$ and the characteristic cluster size in the weighted distribution function $nP\left(n\right)$ does not scale with the particle number $N$. 
In contrast, the dipolar circle swimmers display a macro-flocking state if $\phi_{\boldsymbol{e}} > 0.5$ and the size of the giant clusters corresponding to the peak structure in $nP\left(n\right)$ scales with $N$.
\begin{figure}[!htbp]
  \centering
  \includegraphics[width=0.9\linewidth]{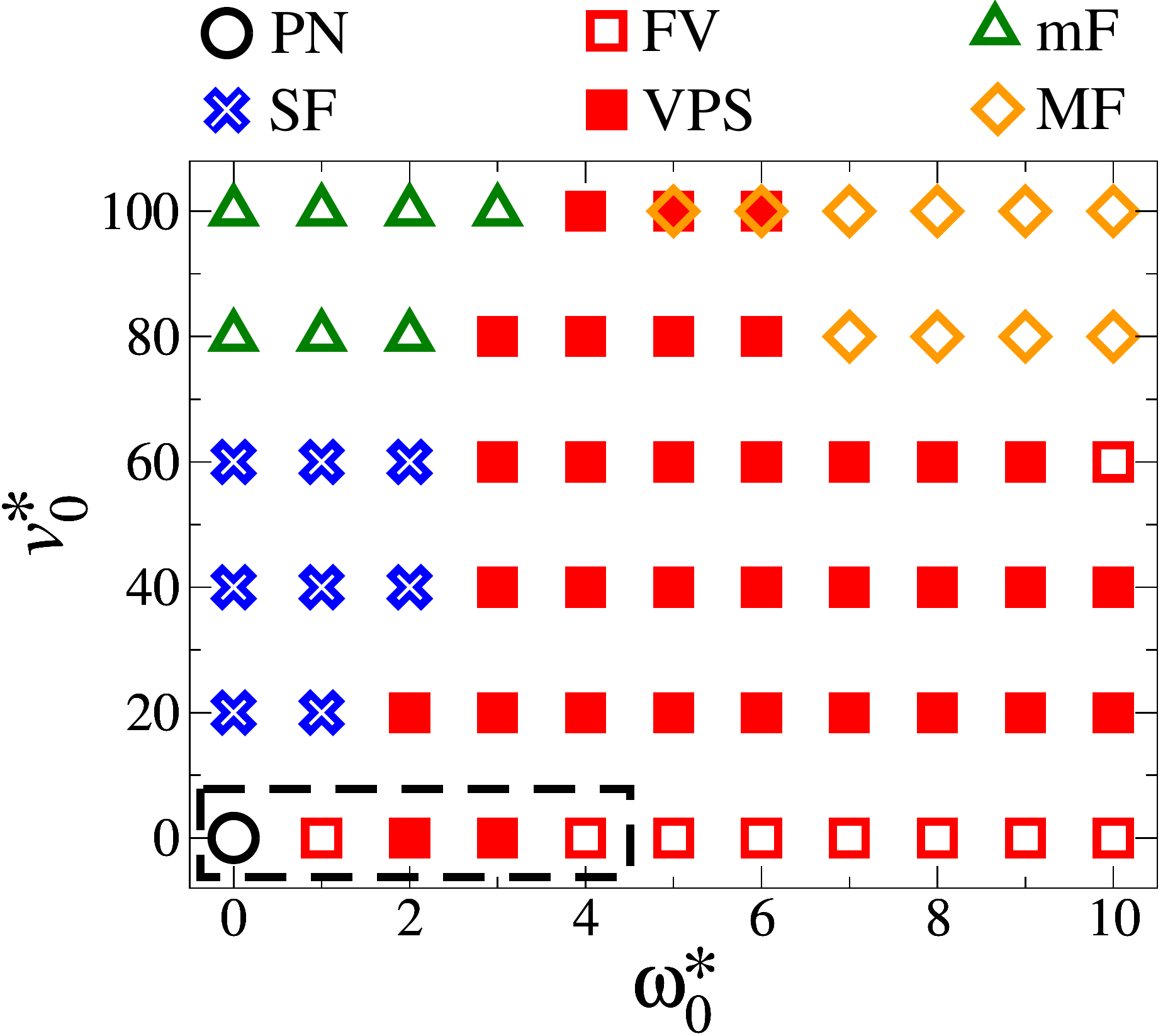}
  \caption{State diagram in the plane spanned by the motility $v_0^*$ and angular speed $\omega_0^*$ at $\Phi = 0.23$. The position of the symbols on the diagram indicates the parameter combinations used in simulations. We have observed percolated networks (PN, black circles), string fluids (SF, blue crosses), finite-size vortices (FV, red open squares), vortices with phase separation (VPS, red filled squares), micro-flocking (mF, green triangles), and macro-flocking (MF, orange diamonds), for details of this classification see Table~\ref{DCSs:table:states}. Overlapping symbols denote bistable states. The region surrounded by the dashed lines indicates a parameter regime where the system is percolated.}
  \label{DCSs:fig:state_diagram}
\end{figure}
\section{Results and discussion}
In this section we present results from extensive BD simulations for a wide range of motilities ($0 \leq v_0^* \leq 100$) and angular speeds ($0 \leq \omega_0^* \leq 10$) at a fixed area fraction $\Phi = 0.23$ and dipolar coupling strength $\lambda = 10$.
In the subsequent subsection~\ref{DCSs:sec:result4}, we first give an overview of the observed states, and then discuss in Sec.~\ref{DCSs:sec:result2}-\ref{DCSs:sec:result3} various aspects in detail.
\begin{figure*}[!htbp]
  \centering
  \includegraphics[width=0.8\linewidth]{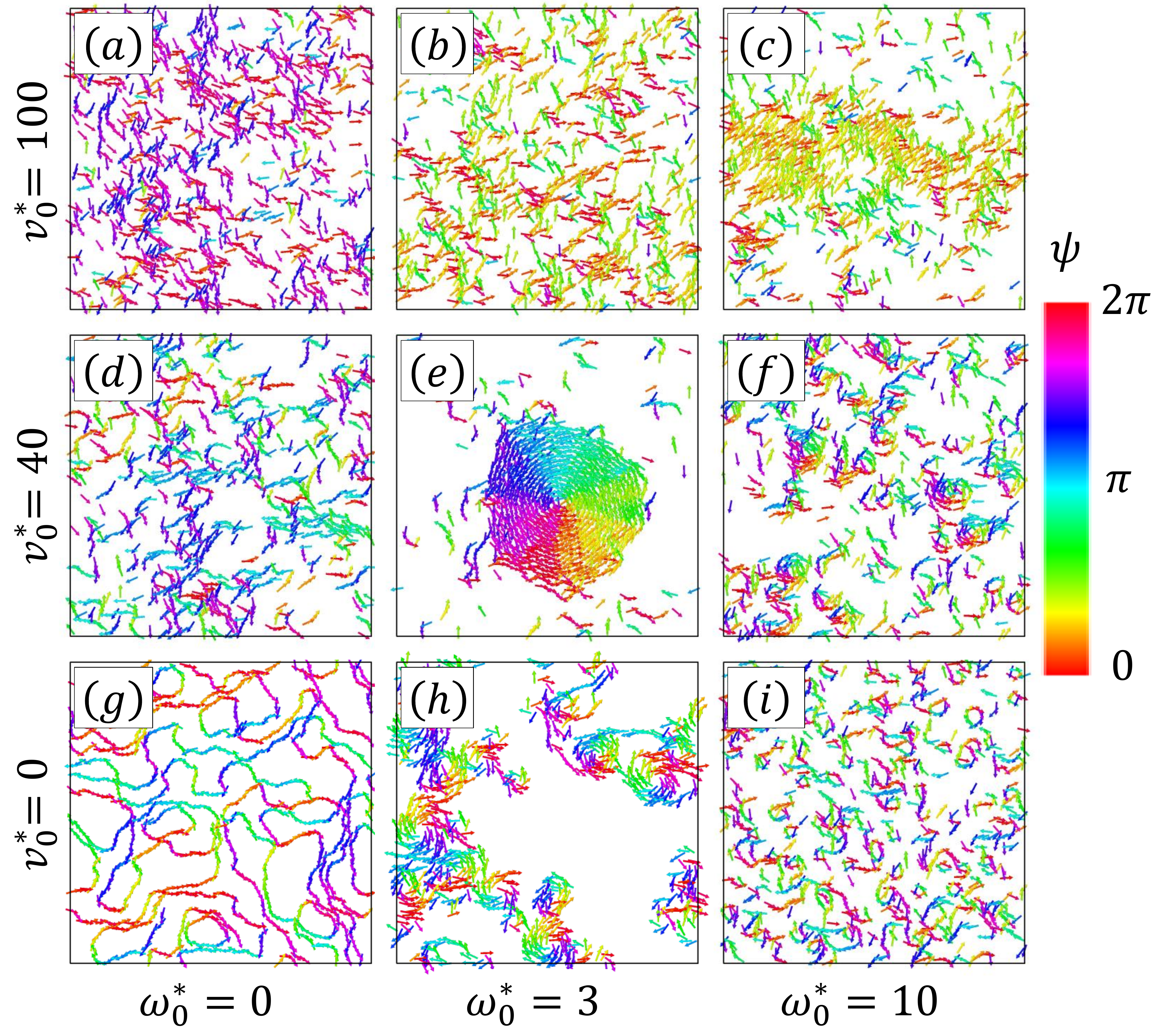}
  \caption{Representative simulation snapshots. Each of the disk-shaped particles is represented by an arrow indicating the particle orientation $\hat{\boldsymbol{e}}_i$, which coincides with the dipole orientation $\hat{\boldsymbol{\mu}}_i$. For better visual quality, all arrows are scaled by a factor of $3$. Colors reflect the direction of particle orientation \textit{via} the angle $\psi_i$ to the $x$-axis. The corresponding steady states include micro-flocking (a, b), macro-flocking (c), string fluids (d), vortices with accompanying phase separation (e, f, h), percolated networks (g), and finite-size vortices (i).}
  \label{DCSs:fig:9snapshots}
\end{figure*}
\subsection{\label{DCSs:sec:result4}Non-equilibrium state diagram}
Based on the targeted quantities described in Sec.~\ref{DCSs:sec:target_quant}, we have identified six states whose characteristics are summarized in Table~\ref{DCSs:table:states}.
The resulting non-equilibrium state diagram spanned in the $\left(v_0^*, \omega_0^*\right)$ plane is shown in Fig.~\ref{DCSs:fig:state_diagram}, and representative simulation snapshots are given in Fig.~\ref{DCSs:fig:9snapshots}.
In the passive limit ($v_0^* = \omega_0^* = 0$), the system displays percolated networks [see Fig.~\ref{DCSs:fig:9snapshots}(g)], which is in agreement with simulation studies for a monolayer of passive dipolar spheres.\cite{Duncan2004, Duncan2006}
For a finite motility $v_0^*$ and in the limit of linear swimmers ($\omega_0^* = 0$), our model reduces to the model of dipolar active Brownian particles.\cite{Liao2020}
Upon increasing $v_0^*$ this system undergoes a transition from a string fluid state characterized by chain-like structures [see Fig.~\ref{DCSs:fig:9snapshots}(d)] into a micro-flocking state with finite-size polar clusters [see Fig.~\ref{DCSs:fig:9snapshots}(a)].
Novel behavior emerges as the angular speed $\omega_0^*$ becomes non-zero.
At vanishing motility ($v_0^* = 0$), a slight increase in $\omega_0^*$ from zero leads to the emergence of two types of vortex patterns, which will be discussed in detail in Sec.~\ref{DCSs:sec:result5}.
As a ``side effect'' of the vortex formation, phase separation occurs at angular speeds $\omega_0^* = 2 - 3$ [see Fig.~\ref{DCSs:fig:9snapshots}(h)].
A further increase in $\omega_0^*$ renders the vortex patterns less pronounced, suppresses the phase separation, and breaks the percolated structures, as shown in Fig.~\ref{DCSs:fig:9snapshots}(i).
Considering a finite motility of $v_0^* = 40$, the swimmers exhibit a string fluid state at small angular speeds ($0 \leq \omega_0^* \lesssim 2$), while intermediate and fast rotation ($3 \lesssim \omega_0^* \lesssim 10$) induce the formation of vortices combined with phase separation, whose sizes decrease with increasing $\omega_0^*$ [see Fig.~\ref{DCSs:fig:9snapshots}(e) and (f)]. 
Finally, at high motilities ($v_0^* \gtrsim 80$) and slow rotation ($\omega_0^* \lesssim 2$), the chain-like structures observed at low motilities break; instead, the system displays a micro-flocking state [see Fig.~\ref{DCSs:fig:9snapshots}(a) and (b)]. 
At intermediate angular speeds ($3 \lesssim \omega_0^* \lesssim 6$), the dipolar circle swimmers self-assemble into vortices with accompanying phase separation.
Here we note that the parameter window of the angular speed $\omega_0^*$ for the vortex formation is narrower than the low motility case ($v_0^* < 80$).
For fast rotation such as $\omega_0^* \gtrsim 7$, the system exhibits a macro-flocking state, where the sizes of flocking patterns are comparable to the simulation box [see Fig.~\ref{DCSs:fig:9snapshots}(c)].
Interestingly, we also observe bistable states involving vortices and macro-flocking at $v_0^* = 100$ and $\omega_0^*= 5 - 6$.
This bistability will be further discussed in Sec.~\ref{DCSs:sec:result2}.

\begin{figure}[!htbp]
  \centering
  \includegraphics[width=0.9\linewidth]{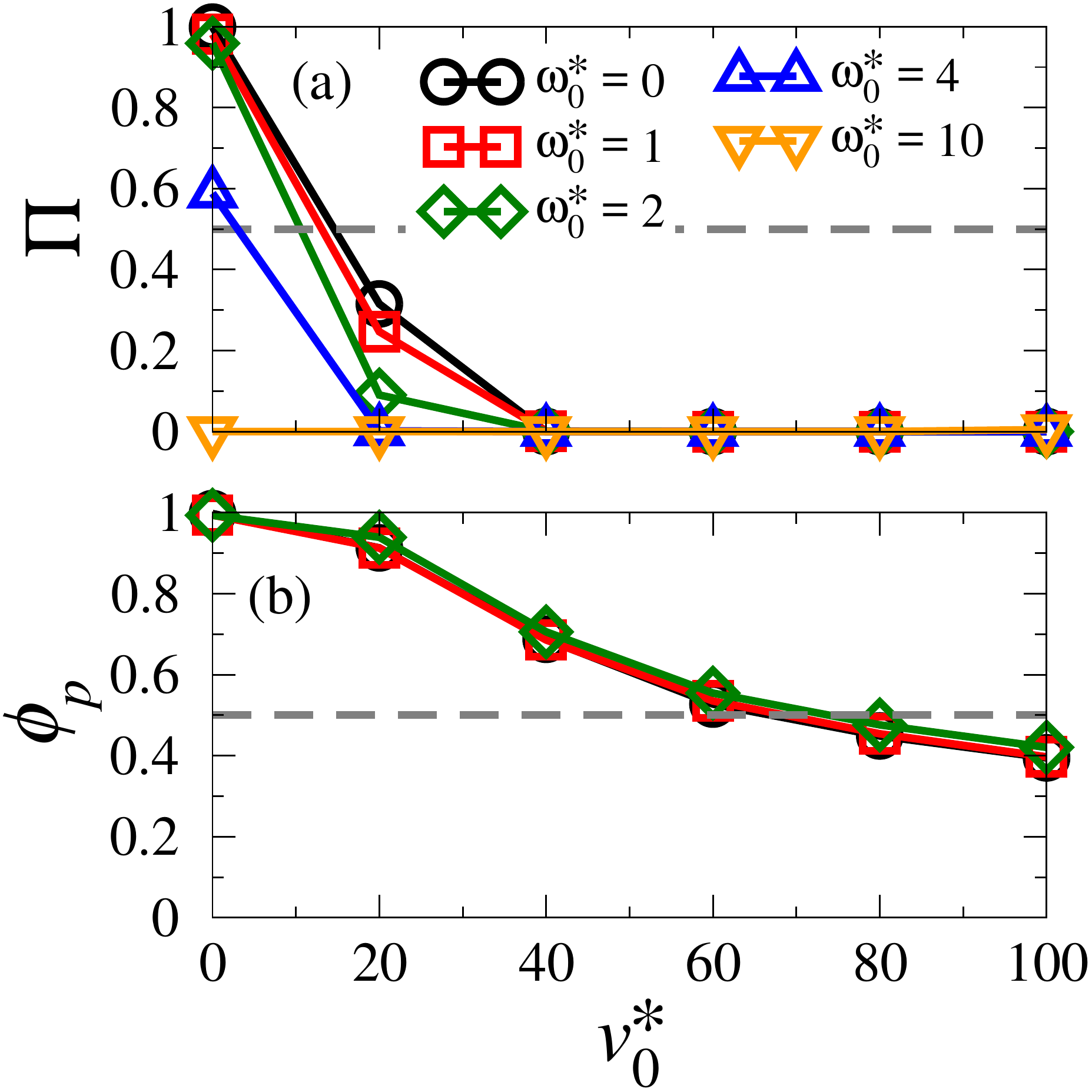}
  \caption{(a) Percolation probability $\Pi$ and (b) degree of polymerization $\phi_p$ as a function of the motility $v_0^*$ for angular speed $\omega_0^* = 0$ (black circles), $1$ (red squares), $2$ (green diamonds), $4$ (blue triangles up), and $10$ (orange triangles down). The dashed horizontal line in each figure marks the values $\Pi = 0.5$ (a) and $\phi_p = 0.5$ (b). Above this line, the system is percolated or polymerized, respectively (see Table~\ref{DCSs:table:states}). The solid lines are guides to the eye.}
  \label{DCSs:fig:percolation}
\end{figure}
\subsection{\label{DCSs:sec:result1}Chain formation}
We start our detailed discussion of the system's behavior by focusing on the bottom left part of the state diagram in Fig.~\ref{DCSs:fig:state_diagram}, where the key phenomenon is chain formation.
We begin by investigating the percolation probability $\Pi$ and the degree of polymerization $\phi_p$ defined in Sec.~\ref{DCSs:sec:chaining}.
Figure~\ref{DCSs:fig:percolation}(a) shows the percolation probability $\Pi$ as a function of the motility $v_0^*$ for various angular speeds $\omega_0^*$.
At $v_0^* = 0$ and $\omega_0^* = 0$, $\Pi$ approaches unity, which corresponds to the percolated networks well known for the passive case.\cite{Laria1991}
Upon an increase of $v_0^*$ from zero at zero angular speed ($\omega_0^* = 0$), $\Pi$ quickly decays and reaches a value close to zero when $v_0^* \gtrsim 40$.
The corresponding snapshot is shown in Fig.~\ref{DCSs:fig:9snapshots}(d).\footnote[2]{
The arrows in Fig.~\ref{DCSs:fig:9snapshots}(d) are chosen to be three times as large as
the diameter of a disk-shaped swimmer for better visualization. 
However, this choice may create an unrealistic visual effect that some neighboring chains in Fig.~\ref{DCSs:fig:9snapshots}(d) seem to be connected and span over the simulation cell.
In fact, it is not the case, since the percolation probability $\Pi$ approaches zero at $v_0^* = 40$ and $\omega_0^* = 0$ [see Fig.~\ref{DCSs:fig:percolation}(a)].
}
For finite (fixed) values of $\omega_0^*$, $\Pi$ decreases more drastically with $v_0^*$, and this decrease becomes the more pronounced the larger $\omega_0^*$ is.
This suggests that both, the motility and the active rotation, tend to suppress the percolation behavior (relative to the passive case).
Once $\Pi$ has dropped to values below $0.5$, the percolated networks have dissolved, yielding a fluid state with chains that do not span over the simulation box.
To quantify these chain structures, we calculate the degree of ``polymerization'' $\phi_p$ defined in Eq.~\eqref{DCSs:eqn:phi_p}.
We recall that $\phi_p$ alone cannot distinguish long string-like chains and compact disk-shaped clusters.
To avoid possible confusion, in the following we only present the results for $\phi_p$ when indeed chain structures are formed.
Results plotted in Fig.~\ref{DCSs:fig:percolation}(b) show that, indeed, the polymerization $\phi_p$ is close to unity at zero motility and gradually decreases as $v_0^*$ increases.
Once $v_0^* \gtrsim 80$, the chain structures essentially disappear.
We also find that the degree of polymerization $\phi_p$ is essentially unaffected by rotation for the small values of $\omega_0^*$ considered here ($\omega_0^* \leq 2$).
The observed decrease of $\phi_p$ at small $\omega_0^*$ (and $\Phi = 0.23$) is, qualitatively, in a good agreement with earlier studies of dipolar active particles, where $\omega_0^* = 0$.\cite{Liao2020}
Once $\omega_0^* > 2$, the dipolar circle swimmers may self-assemble into patterns distinct from chain structures, such as compact clusters, which we will discuss in the following sections.
\begin{figure}[!htbp]
  \centering
  \includegraphics[width=0.95\linewidth]{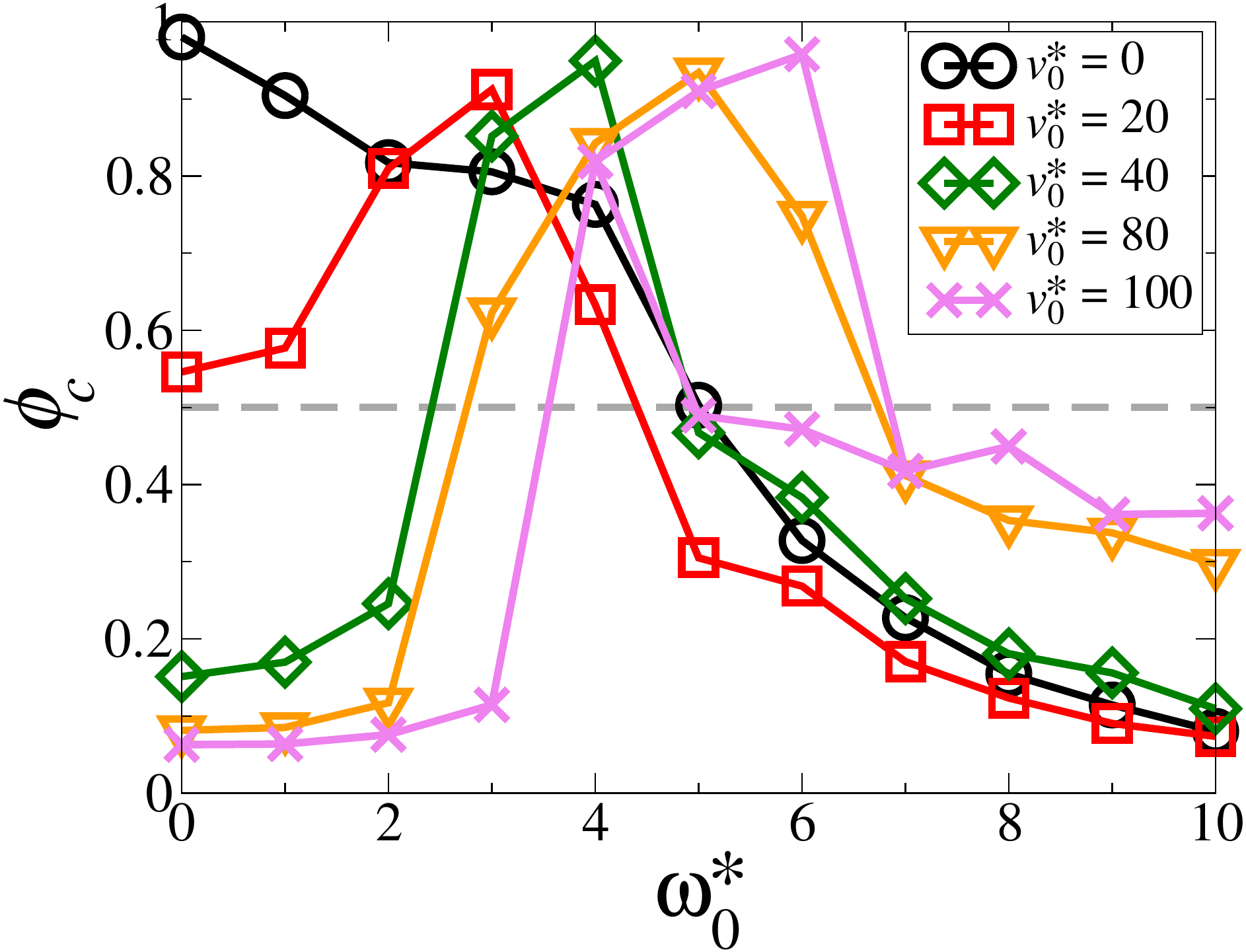}
  \caption{Fraction of the largest cluster $\phi_c$ as a function of the angular speed $\omega_0^*$ for the motility $v_0^* = 0$ (black circles), $20$ (red squares), $40$ (green diamonds), $80$ (orange triangles down), and $100$ (violet crosses). At $v_0^* = 100$ and $\omega_0^* \approx 5 - 6$, there is a bistability as indicated by the two values of $\phi_c$, see Sec.~\ref{DCSs:sec:result2} for more details. The dashed line marks the criterion $\phi_c = 0.5$, above which the system displays giant clusters.}
  \label{DCSs:fig:phi_c}
\end{figure}
\subsection{\label{DCSs:sec:result2}Clustering and phase separation}
It is well established that non-dipolar active Brownian particles can exhibit clustering and even phase separation at sufficiently large motilities (and appropriate densities).\cite{Buttinoni2013, Speck2014, Speck2015, Liao2018, Sese-Sansa2018}
In the absence of active rotation or alignment interactions,\cite{Buttinoni2013, Speck2014, Speck2015, Cates2015} the clustering behavior is purely induced by the interplay between the particle motility and the steric repulsion.
To quantify the clustering behavior in the present system, we first calculate the fraction of the largest cluster $\phi_c$, see Eq.~\eqref{DCSs:eqn:phi_c} and Fig.~\ref{DCSs:fig:phi_c}.
In the passive limit ($v_0^* = 0$ and $\omega_0^* = 0$), one finds $\phi_c \approx 1$.
This is due to the large value of $\lambda$, which leads to strong head-tail-alignment and thus, to the formation of percolated networks, with percolation probability $\Pi \approx 1$ [see Fig.~\ref{DCSs:fig:9snapshots}(g) and Fig.~\ref{DCSs:fig:percolation}(a)].
Upon an increase of $\omega_0^*$ from zero at $v_0^* = 0$, the order parameter $\phi_c$ decreases (as does $\Pi$), indicating that the percolated networks are suppressed by the active rotation.
This seems plausible, since the active rotation tends to destroy the head-to-tail alignment.
This leads to a breaking of percolated chains and, hence, a decrease of $\phi_c$.
At finite $v_0^*$ the situation changes.
Inspecting Fig.~\ref{DCSs:fig:percolation}(a) again, we see that at $v_0^* = 20$, the system is only partially percolated (with $\Pi \lesssim 0.5$) for $\omega_0^* \lesssim 2$, and if $v_0^* \gtrsim 40$, the system is not percolated at all (with $\Pi \approx 0$) for all explored $\omega_0^*$.
We infer from the data that for $v_0^* \gtrsim 20$, the values of $\phi_c > 0.5$ seen in Fig.~\ref{DCSs:fig:phi_c} truly indicate the formation of giant compact clusters, rather than that of percolated networks.
We also find from Fig.~\ref{DCSs:fig:phi_c} that for $v_0^* \gtrsim 20$, $\phi_c$ varies non-monotonically with $\omega_0^*$.
We interpret this interesting observation such that giant clusters only appear at intermediate angular speeds ($2 \lesssim \omega_0^* \lesssim 6$), whereas the system is rather homogeneous ($\phi_c \leq 0.5$) for slow and fast rotation. 
Further, with increasing motilities $v_0^*$, the window of $\omega_0^*$ values corresponding to giant clusters is shifted toward larger $\omega_0^*$.
We conclude that at finite $\omega_0^*$, the motility tends to suppress (rather than enhance) the formation of giant clusters.
\begin{figure}[!hbtp]
  \centering
  \includegraphics[width=0.9\linewidth]{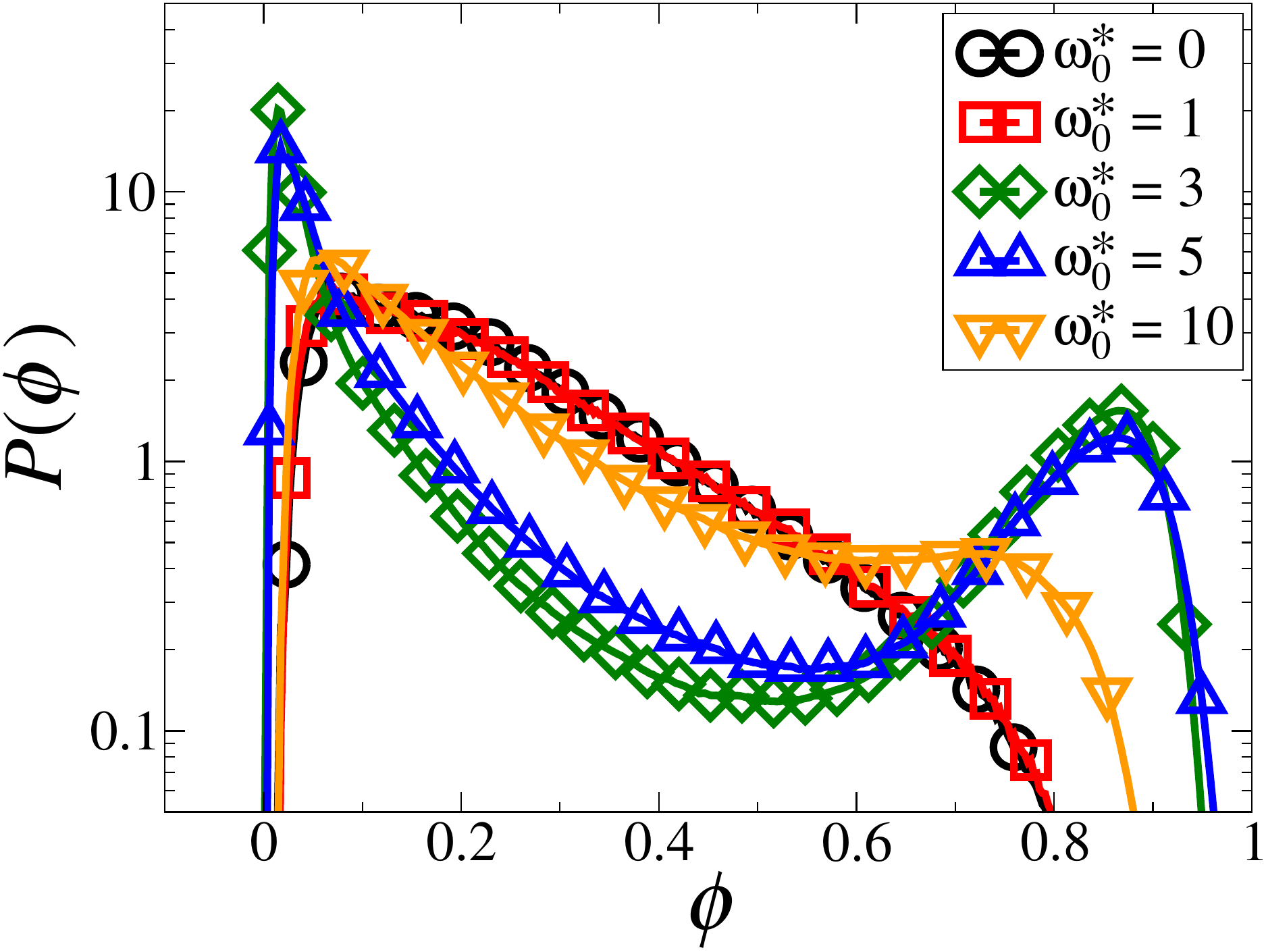}
  \caption{Probability distribution function of the local area fraction $P(\phi)$ for the angular speed $\omega_0^* = 0$ (black circles), $1$ (red squares), $3$ (green diamonds), $5$ (blue triangles up), and $10$ (orange triangles down) at the motility $v_0^* = 40$.}
  \label{DCSs:fig:P_of_phi}
\end{figure}
A particularly complex situation occurs at very high motilities.
For example, at $v_0^* = 100$ we observed that the system of dipolar circle swimmers can change within a single simulation from one state with $\phi_c \approx 1$ (indicating giant clusters) to another state with $\phi_c \lesssim 0.5$, or the other way around (snapshots not shown here). 
To check whether both of these states are steady states, we created multiple long-time realizations at $v_0^* = 100$ and various $\omega_0^*$.
Indeed, two different steady-state results were found from independent simulation realizations in the range $\omega_0^* = 5 - 6$, indicating a bistability.
Specifically, the realizations with $\phi_c \approx 1$ are characterized by giant clusters, while those with $\phi_c \lesssim 0.5$ correspond to a macro-flocking state, where macroscopic swarming is observed.
The flocking behavior of dipolar circle swimmers will be discussed later in detail in Sec.~\ref{DCSs:sec:result3}.
Given the formation of giant clusters at suitable combinations of $v_0^*$ and $\omega_0^*$, it is interesting to explore whether this leads to phase separation.
To this end, we plot in Fig.~\ref{DCSs:fig:P_of_phi} the probability distribution function $P(\phi)$ of the local area fraction $\phi$ for various angular speeds $\omega_0^*$, taking the motility $v_0^* = 40$ as an example.
Phase separation is indicated by a double-peak structure of $P(\phi)$, where the coexisting densities are the density values corresponding to the peaks. 
From Fig.~\ref{DCSs:fig:P_of_phi} we see that at zero and small angular speeds ($\omega_0^* \lesssim 1$), the distribution function displays only a single peak located at $\phi \approx 0.1$, indicating that the system is essentially homogeneous.
It is noted that at $v_0^* = 40$ and $\omega_0^* = 0$ or $1$, the system exhibits a state with string fluids (see Fig.~1), \textit{i.e.}, particles tend to form short linear chains. 
As a result, the distribution function is not symmetric and shows a pronounced tail at high densities.
In contrast, at intermediate angular speeds ($3 \lesssim \omega_0^* \lesssim 9$), we observe that $P\left(\phi\right)$ exhibits two well-defined peaks, showing that the dipolar circle swimmers phase-separate into dilute and dense domains.
Finally, at a large angular speed $\omega_0^* = 10$, the second maximum at high densities is only weakly pronounced, indicating that the dense domains almost disappear.
Therefore, we expect that phase separation will eventually vanish upon further increase of $\omega_0^*$.

\begin{figure}[!htbp]
  \centering
  \includegraphics[width=0.9\linewidth]{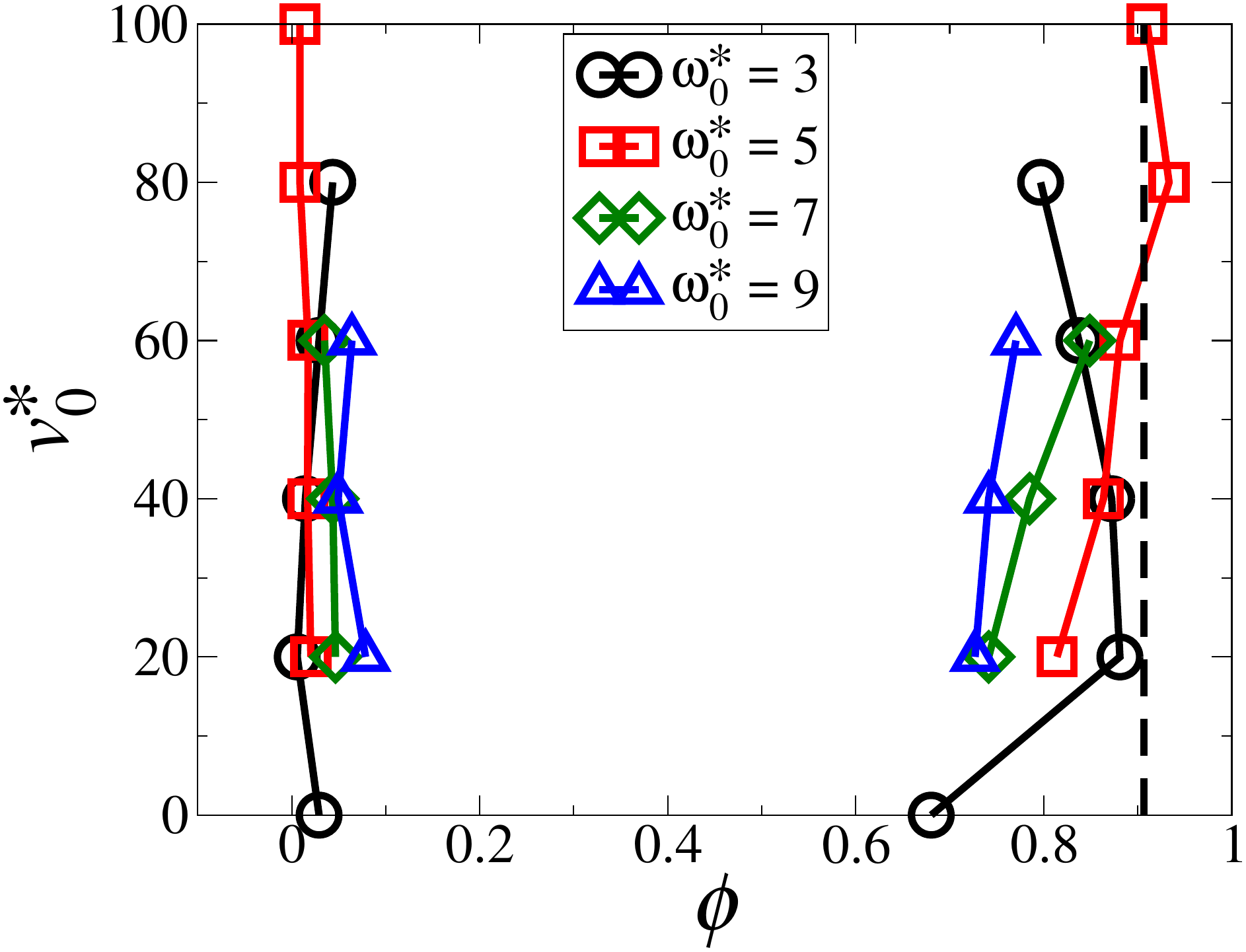}
  \caption{Densities of the coexisting states in the ($v_0^*$, $\phi$) plane for the angular speed $\omega_0^* = 3$ (black circles), $5$ (red squares), $7$ (green diamonds), and $9$ (blue triangles up). The black dashed line marks the close-packing fraction, $\phi_{cp} = \phi / \sqrt{12} \approx 0.91$.}
  \label{DCSs:fig:binodal}
\end{figure}
Finally, we plot in Fig.~\ref{DCSs:fig:binodal} the coexistence densities in the ($v_0^*$, $\phi$) plane for various angular speeds $\omega_0^*$.
At an intermediate angular speed $\omega_0^* = 3$, the system displays phase separation for a broad range of the motility, $v_0^* \approx 0 - 80$.
Further, the density difference, $\Delta_{\phi} = \phi_d - \phi_g$, between the dense and the gas-like region first increases with $v_0^*$ from zero to $20$, and then decreases upon further increasing $v_0^*$ from $20$ to $80$.
This non-monotonic behavior of $\Delta_{\phi}$ may be attributed to the fact that the parameter regime for $\omega_0^* = 3$ and $v_0^* = 20 - 80$ is very close to the state boundary between string fluids and vortices with phase separation (see Fig.~\ref{DCSs:fig:state_diagram}).
As $\omega_0^*$ increases from $3$ to $5$, the motility range where phase separation occurs is shifted toward larger motilities ($v_0^* = 20 - 100$) with $\Delta_{\phi}$ increasing monotonically with $v_0^*$.
At large angular speeds such as $\omega_0^* \gtrsim 7$, the range of the motility shrinks to $30 \lesssim v_0^* \lesssim 50$.
Moreover, the area enclosed by the curves of coexisting densities decreases with $\omega_0^*$, suggesting that phase separation is in general suppressed by $\omega_0^*$.
The suppression of phase separation is also consistent with the vanishing peak of $P(\phi)$ at high densities for $\omega_0 = 10$ in Fig.~\ref{DCSs:fig:P_of_phi}.
Indeed, at extremely large $\omega_0^*$ each of the particles tends to swim along a small circular path.
Thus, it quickly alters its propulsion direction, causing the large dense domain to ``melt'' and break into small pieces.
Similar observations regarding the impact of $\omega_0^*$ on phase separation have been made for systems of non-dipolar circle swimmers.\cite{Liao2018}

\subsection{\label{DCSs:sec:result5}Emergent vortices}
In this section we explore in detail an intriguing ``byproduct'' of the phase separation discussed in Sec.~\ref{DCSs:sec:result2}, namely, the formation vortices occupying the dense domain [for an illustration, see Fig.~\ref{DCSs:fig:9snapshots}(e)]. 
\subsubsection{\label{DCSs:sec:vortexSize}Vortex types and sizes}
\begin{figure}[!htbp]
  \centering
  \includegraphics[width=1.0\linewidth]{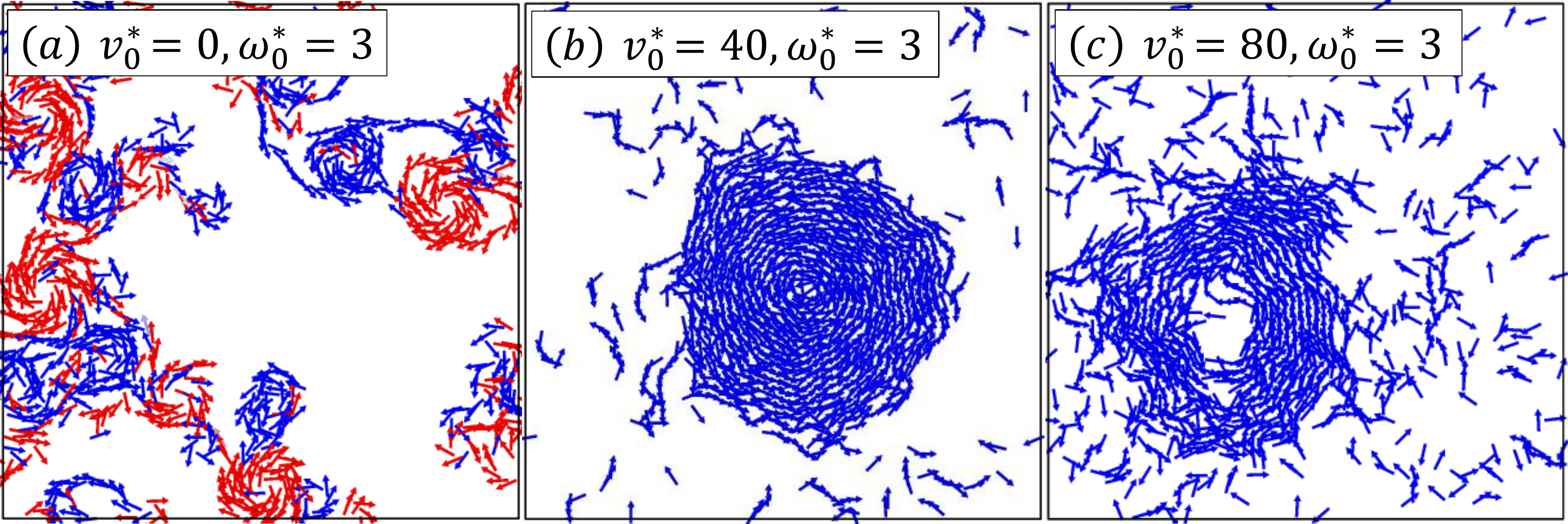}
  \caption{Representative simulation snapshots at $\omega_0^* = 3$ and three values of the motility. Each of the disk-shaped particles is represented by an arrow indicating the particle orientation $\hat{\boldsymbol{e}}_i$. For better visual quality, all arrows are scaled by a factor of $3$ and are formatted with red (blue) color if the particle moves against (along) its own orientation [see Eq.~\eqref{DCSs:eqn:direction_of_motion}]. The blue arrows display Type I vortices (a-c), and the red arrows exhibit Type II vortices (a). }
  \label{DCSs:fig:3snapshots}
\end{figure}

We start our analysis of vortex structures by reconsidering the simulation snapshots in Fig.~\ref{DCSs:fig:9snapshots}(h) and (e) with an alternative coloring scheme illustrating the motion of each particle, as shown in Fig.~\ref{DCSs:fig:3snapshots}.
Specifically, we characterize the motion of particle $i$ relative to its orientation $\hat{\boldsymbol{e}}_i = \hat{\boldsymbol{\mu}}_i$ by a parameter $g_i\left(t\right)$, defined as
\begin{equation}
   g_i\left(t\right) = 
     \boldsymbol{v}_{i,s}\left(t\right)
     \cdot
     \hat{\boldsymbol{e}}_i\left(t\right) \text{,} 
   \label{DCSs:eqn:direction_of_motion}
\end{equation}
where the (average) velocity $\boldsymbol{v}_{i,s}\left(t\right)$ for particle $i$ in a short time interval between $t$ and $t + \Delta t_s$ is given by
\begin{equation} \label{DCSs:eqn:vel}
   \boldsymbol{v}_{i,s}\left(t\right) = 
     \dfrac{
       \boldsymbol{r}_{i}\left(t+\Delta t_s\right) 
       - 
       \boldsymbol{r}_{i}\left(t\right) 
     }
     {
       \Delta t_s
     } \text{.}
\end{equation}
The arrow corresponding to particle $i$ in Fig.~\ref{DCSs:fig:3snapshots} is colored in red if $g_i\left(t\right) \leq 0$, meaning that the particle moves ``backward'' (against its orientation).
In contrast, the arrow is colored in blue if the particle stops or moves ``forward'' (along its orientation). 
When calculating $\boldsymbol{v}_{i,s}$, we take into account the fact that an active rotating particle with zero motility needs more time to move a distance equal to its diameter than a highly motile circle swimmer. 
Therefore, we make the choice $\Delta t_s = \tau$ for the case of $v_0^* = 0$ and $\Delta t_s = 0.01\tau$ for $v_0^* \geq 40$.
All color-coded snapshots in Fig.~\ref{DCSs:fig:3snapshots} refer to $\omega_0^* = 3$.
At zero motility ($v_0^* = 0$), the particles have no tendency to move along their orientation.
Therefore, the numbers of particles moving forward (blue) and backward (red) are on average equal, consistent with Fig.~\ref{DCSs:fig:3snapshots}(a).
A surprising result, however, is that forward- and backward-moving particles \textit{aggregate} with particles displaying the same kind of motion, yielding vortices.
In each instant of time, the chirality of the vortex structure formed by the forward-moving particles is counterclockwise, which we refer to as Type I vortices. 
In contrast, the chiral structure of the backward-moving particles is clockwise (Type II vortex).
We note, however, that since the intrinsic active rotation drives each of the particles to rotate counterclockwise (see Sec.~\ref{DCSs:sec:EoM}), the direction of overall rotation for both, Type I and Type II vortices, is counterclockwise.
The detailed discussion about the physical origin of the emergent Type I and Type II vortices is postponed to Sec.~\ref{DCSs:sec:ring}.
Once the motility becomes non-zero, the particles tend to self-propel along their orientation (instantaneous) orientation.
We thus expect that values of $g_i > 0$ become more and more relevant.
Indeed, as seen from Fig.~\ref{DCSs:fig:3snapshots}(b), we only observe forward-moving swimmers which are represented by blue arrows as $v_0^*$ increases from zero to $40$.
Comparing Fig.~\ref{DCSs:fig:3snapshots}(a) and (b), we further see that the vortex size significantly increases with $v_0^*$.
At an even higher motility [$v_0^* = 80$, see Fig.~\ref{DCSs:fig:3snapshots}(c)], the disk-shaped vortex structures seen at lower motilities are slightly distorted and have a hole in the vortex center.
To understand this, we recall that an isolated circle swimmer moves along a circular path with a radius $R = v_0 / \omega_0$.\cite{VanTeeffelen2008}
Thus, $R$ increases with $v_0$.
This ideal motion, however, is disturbed by the presence of other swimmers, which makes it increasingly difficult to stay in the center of a giant vortex when $R$ becomes larger.
This eventually leads to a hole in the vortex center.

\begin{figure}[!htbp]
  \centering
  \includegraphics[width=0.9\linewidth]{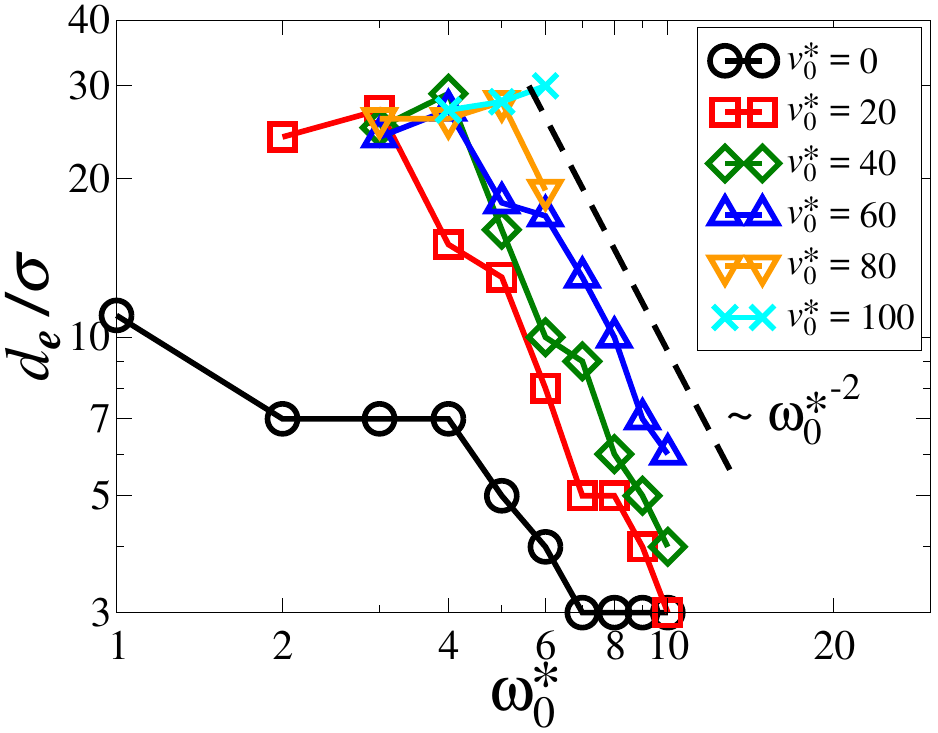}
  \caption{The vortex diameter, $d_e$, as a function of the angular speed $\omega_0^*$ for the motility $v_0^* = 0$ (black circles), $20$ (red squares), $40$ (green diamonds), $60$ (blue triangles up), $80$ (orange triangles down), and $100$ (cyan crosses). The dashed line indicates the scaling behavior.}
  \label{DCSs:fig:d_e}
\end{figure}
Further information on the vortex structure and size is provided by the orientational correlation function $C_{\boldsymbol{e}}(r)$, whose detailed analysis is presented in Appendix~\ref{DCSs:sec:Ce_of_r}.
Here, we focus on the vortex diameter $d_{\boldsymbol{e}}$, which is obtained from the minimum of $C_{\boldsymbol{e}}(r)$ (see Sec.~\ref{DCSs:sec:vortex}).
Figure~\ref{DCSs:fig:d_e} shows $d_{\boldsymbol{e}}$ as a function of the angular speed $\omega_0^*$ for various motilities $v_0^*$.
Upon an increase of $v_0^*$, the range of $\omega_0^*$ where vortices appear is narrowed.
For all explored motilities, $d_{\boldsymbol{e}}$ generally decreases with the angular speed $\omega_0^*$, and the curves are shifted toward a larger $d_{\boldsymbol{e}}$ as $v_0^*$ increases. 
Specifically, we observe a power law decay with an exponent $\nu \approx -2$ for $v_0^* \gtrsim 20$.
This exponent is different from that observed in studies of the rotating Vicsek model (and variants) where $\nu = -1$.\cite{Liebchen2017, Levis2018}
Finally, at a fixed motility such as $v_0^* = 40$ and upon an increase of the angular speed from $\omega_0^* = 0$ to $\omega_0^* = 3$, the system undergoes a transition from a state with short linear chains (\textit{i.e.}, string fluids) to a state with giant vortices, as shown in Fig.~2(d) and 2(e). 
Here we identify the transition by a drastic increase of $\phi_c$ (see Fig.~4), and by the appearance of negative orientational correlations [see Fig.~12(b)]. 
Clearly, it would be very interesting to see whether this non-equilibrium transition is/resembles a first-order or continuous transition. 
However, such a study would require more extensive simulations and a full analysis of the order parameters, which is out of scope for the present paper.
\begin{figure}[!htbp]
  \centering
  \includegraphics[width=1.0\linewidth]{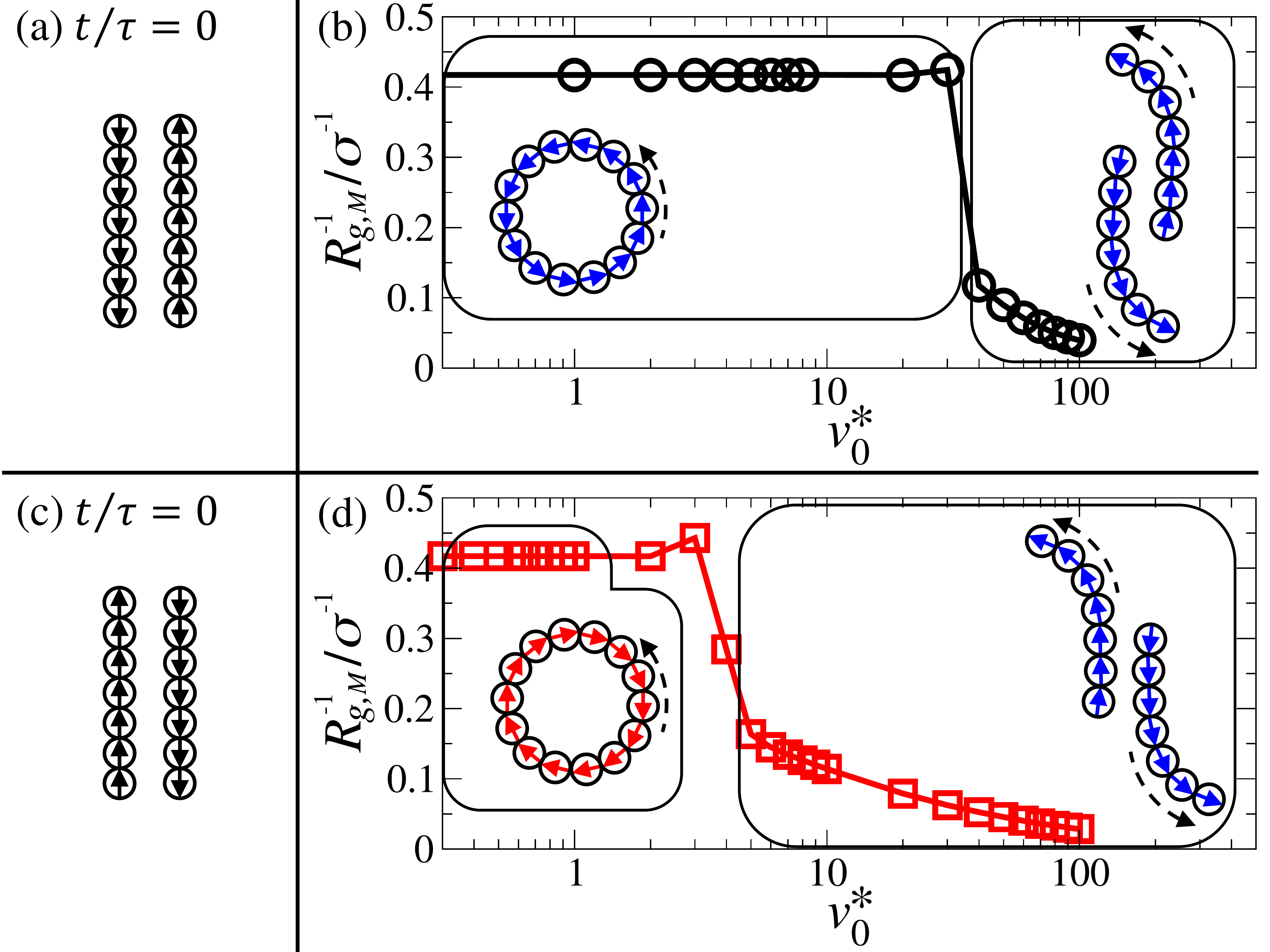}
  \caption{(a, c) Sketches of two initial chain configurations composed of $N_t = 14$ dipolar circle swimmers. The black arrows indicate the particle orientation at $t / \tau = 0$. (b, d) The inverse of maximum gyration radius, $R_{g, M}^{-1}$, between $30 \leq t / \tau \leq 40$ as a function of the motility $v_0^*$ for initial configuration (a) (black circles) and (c) (red squares) at the angular speed $\omega_0^* = 3$. Sketches show respective configurations characterized by $R_{g, M}^{-1}$, where the red (blue) arrow indicates that the particle's velocity points against (along) its orientation [see Eq.~\eqref{DCSs:eqn:direction_of_motion} with $\Delta t_s = 0.01 \tau$] and the dashed arrows indicate the particle motion. The rectangles with curved angles relate data with corresponding sketches.}
  \label{DCSs:fig:sketch}
\end{figure}
\subsubsection{\label{DCSs:sec:ring}Ring argument}
To understand the physical origin of the emergent vortices in the present system, we first recall that passive dipolar particles, at the large coupling strength considered here, self-assemble into chains, see Fig.~\ref{DCSs:fig:9snapshots}(g).
Moreover, particles in neighboring chains arranged side-by-side tend to point in anti-parallel directions.
Inspired by this behavior we performed, at an intermediate angular speed $\omega_0^* = 3$, test simulations of a simplified double-chain structure to investigate the impact of the motility $v_0^*$.
As an initial configuration, $N_t$ dipolar particles are arranged into two anti-parallel chains lying side by side, see Fig.~\ref{DCSs:fig:sketch}(a) and~(c). 
Within each chain, the dipole moments are oriented head-to-tail.
As illustrated in Fig.~\ref{DCSs:fig:sketch}(a) and~(c), we consider two different arrangements of these double-chain configurations, which are the mirror image of each other, meaning that they have opposite ``chirality.''
The center-to-center distance between the two chains is chosen to be $2\sigma$. 
Each of the chains is composed of $N_t / 2 = 7$ dipolar particles, such that the resulting chain length corresponds to the vortex diameter $d_{\boldsymbol{e}} / \sigma \approx 7$ observed at $v_0^* = 0$ and $\omega_0^* = 3$ (see Fig.~\ref{DCSs:fig:d_e}).
For simplicity, the thermal fluctuations are neglected.
%

%% New Paragraph
%
Upon starting the simulations with $\omega_0^* = 3$ and $v_0^* = 0$ we observe, for both initial configurations, that the swimmers rearrange themselves into a stable ring structure.
The difference, however, is that the particles in the ring move forward (like in a Type I vortex) when starting from configuration (a), while the particles move backward when starting from the other initial configuration.
This difference is due to the competition between the ring chirality favored by the initial configuration and the counterclockwise rotation supported by the active rotation.
To further characterize the ring structure, we consider the gyration radius, given by
\begin{equation}
  R_g\left(t\right) = \sqrt{\dfrac{1}{2N_t^2}\sum_{i = 1}^{N_t}\sum_{j = 1}^{N_t}\left(\boldsymbol{r}_i - \boldsymbol{r}_j\right)^2}\text{.}
\end{equation}
The radius $R_g\left(t\right)$ is a constant once the chains form a stable ring, while $R_g\left(t\right)$ varies with time $t$ if the chains remain separate and do not self-assemble into a ring even after a long time.
With this in mind, we plot in Fig.~\ref{DCSs:fig:sketch}(b) and~(d) the inverse of the maximum gyration radius $R_{g,M}^{-1}/\sigma^{-1}$ within a time interval $30 \leq t / \tau \leq 40$ as a function of the motility $v_0^*$ for the initial configuration sketched in Fig.~\ref{DCSs:fig:sketch}(a) and~(c), respectively.
The choice of this time interval is based on the fact that no transient structure is observed at $t / \tau > 20$.
The quantity $R_{g,M}^{-1}/\sigma^{-1}$ is about $0.4$ for a stable ring structure and reaches a value smaller than $0.1$ if two chains do not form a ring.
As $v_0^* \ll 1$, both Type I and Type II rings are stable, which conforms with the emergence of Type I and Type II vortices at $v_0^* = 0$ and $\omega_0^* = 3$ [see Fig.~\ref{DCSs:fig:9snapshots}(h) and Fig.~\ref{DCSs:fig:3snapshots}(a)].
We also see that two chains no longer form a ring once $v_0^*$ is greater than a ``critical'' motility $v_{0,c}^*$.
In this latter case, each chain swims along a circular path and does not collide with the other chain within the simulation time, see the sketches on the right side of Fig.~\ref{DCSs:fig:sketch}(b) and~(d).
Importantly, $v_{0,c}^*$ for the Type II ring is much smaller ($v_{0,c}^* \approx 3$) than that for the Type I ring ($v_{0,c}^* \approx 30$).
This explains why the system is dominated by the Type I vortices at intermediate motilities [see Fig.~\ref{DCSs:fig:9snapshots}(e) and~\ref{DCSs:fig:3snapshots}(b)].
Further increasing $v_0^*$ leads to the breaking of both Type I and Type II rings [see Fig.~\ref{DCSs:fig:9snapshots}(b)], suggesting that vortices will eventually vanish for large $v_0^*$.
Indeed, for $v_0^* \gtrsim 80$ the system can display flocking behavior, which we will further address in the following Sec.~\ref{DCSs:sec:result3}.
To check whether the vortex emergence is sensitive to the system size of the simulations, we performed various test simulations at $v_0^* = 40$ and $\omega_0^* = 4$ with the particle number ranging from $N = 100$ to $N = 2500$. 
We observed that giant vortex structures already emerge at $N \approx 400$ and persist up to $N = 2500$ (results not shown here). 
The vortex formation in the present work is of fundamental difference from the emergence of the clockwise vortices in systems of simple (non-dipolar, disk-shaped) circle swimmers,\cite{Liao2018} and the macroscopic droplets reported in studies of the rotating Vicsek model\cite{Liebchen2017} and its variant.\cite{Levis2018}
For simple circle swimmers, the clockwise vortices do not appear at a density lower than the critical density of MIPS, $\Phi_{crit} \approx 0.3$.\cite{Liao2018}
Further, their formation relies on the steric collision of particles in the dilute region with the boundary of giant clusters.
In contrast, for dipolar circle swimmers, the vortex patterns appear already at densities lower than $\Phi_{crit}$.
For circle swimmers with polar alignment, such as the rotating Vicsek model\cite{Liebchen2017} and its variant,\cite{Levis2018} macroscopic droplets emerge at small angular speed $\omega_0^*$.
Inside the droplets, particles align themselves with their neighbors and are directed along a single direction.
This direction rotates in response to the active rotation exerted on each of the particles.
These macroscopic droplets are significantly different from the giant vortices observed in dipolar circle swimmers, suggesting that the type of alignment interactions is crucial for the pattern formation of circle swimmers.
\begin{figure}[!htbp]
  \centering
  \includegraphics[width=0.9\linewidth]{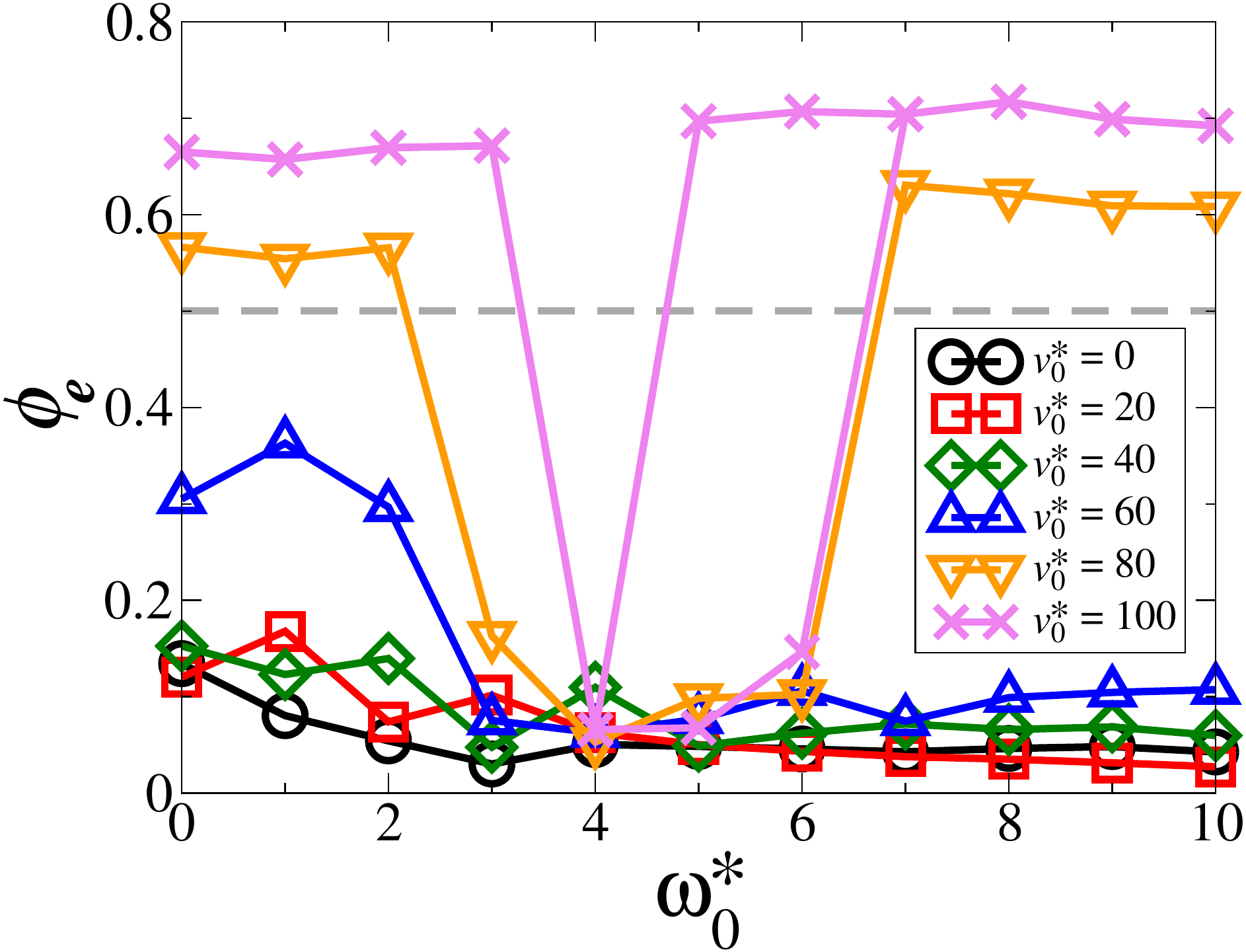}
  \caption{Global polarization $\phi_{\boldsymbol{e}}$ as a function of the angular speed $\omega_0^*$ for the motility $v_0^* = 0$ (black circles), $20$ (red squares), $40$ (green diamonds), $60$ (blue triangles up), $80$ (orange triangles down), and $100$ (violet crosses). At $v_0^* = 100$ and $\omega_0^* = 5 - 6$, the two values of $\phi_{\boldsymbol{e}}$ reflect the bistability already indicated in Fig.~\ref{DCSs:fig:phi_c}. The dashed line marks where $\phi_{\boldsymbol{e}} = 0.5$. The results that lie above this line indicate that the system is in a flocking state.}
  \label{DCSs:fig:phi_e}
\end{figure}
\subsection{\label{DCSs:sec:result3}Orientational ordering}
In this last section, we switch our focus onto motility-induced orientational ordering appearing at large $v_0^*$ (see the top part of the state diagram in Fig.~\ref{DCSs:fig:state_diagram}).
We note that in the present model, the orientational order of dipole moments $\boldsymbol{\mu}_i$ implies coherent motion, \textit{i.e.}, flocking.
The orientational order can be characterized by the global polar order parameter $\phi_{\boldsymbol{e}}$, defined in Eq.~\eqref{DCSs:eqn:phi_e}.
Figure~\ref{DCSs:fig:phi_e} shows $\phi_{\boldsymbol{e}}$ as a function of the angular speed $\omega_0^*$ for various motilities $v_0^*$.
At low motilities ($v_0^* \lesssim 60$), $\phi_{\boldsymbol{e}}$ decreases monotonically with $\omega_0^*$, and there is no significant ordering behavior. 
In contrast, strong translational self-propulsion ($v_0^* \gtrsim 80$) induces significant global polarization at small and large angular speeds ($\omega_0^* \lesssim 2$ and $\omega_0^* \gtrsim 7$) [see Fig.~\ref{DCSs:fig:9snapshots}(a)-(c)], while at intermediate angular speeds ($3 \lesssim \omega_0^* \lesssim 6$) we find that $\phi_{\boldsymbol{e}} \leq 0.2$. Here, the system forms vortex patterns [see Fig.~\ref{DCSs:fig:d_e} and Table~\ref{DCSs:table:states}].
Further, Fig.~\ref{DCSs:fig:phi_e} reveals that the system reaches bistable states at $v_0^* = 100$ and $\omega_0^* = 5 - 6$, consistent with earlier discussion of Fig.~\ref{DCSs:fig:phi_c}.

\begin{figure}[!htbp]
  \centering
  \includegraphics[width=0.9\linewidth]{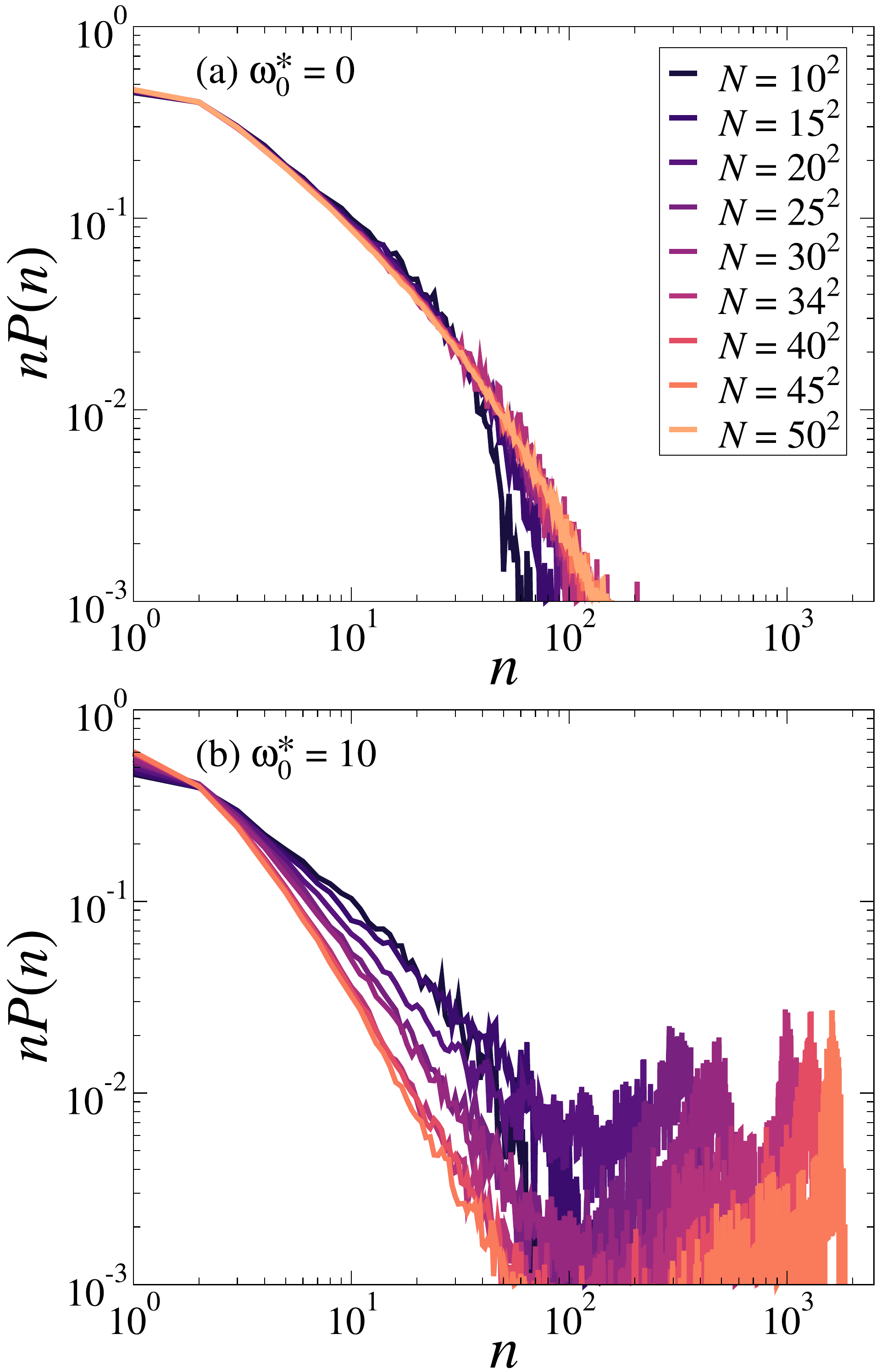}
  \caption{Weighted distribution of cluster size at $v_0^* = 100$ and $\omega_0^* = 0$ (a) and $\omega_0^* = 10$ (b) for various particle numbers $N$.}
  \label{DCSs:fig:nPn}
\end{figure}

Taking a closer look at Fig.~\ref{DCSs:fig:9snapshots}(a)-(c), we find that the size of the ordered (``flocking'') structures formed at large $v_0^*$ significantly depends on $\omega_0^*$.
To further characterize this behavior we consider the (weighted) distribution of the cluster size, $nP\left(n\right)$, and their dependence on the overall particle number $N$.
Results for $v_0^* = 100$ are shown in Fig.~\ref{DCSs:fig:nPn}. 
As can bee seen in Fig.~\ref{DCSs:fig:nPn}(a), for zero rotation ($\omega_0^* = 0$) the distributions decrease monotonically with the cluster size $n$ and collapse onto one curve for $N \gtrsim 900$. 
This indicates that these are only small clusters whose size does not scale with the particle number $N$.
According to Table~\ref{DCSs:table:states}, we classify such a situation as a ``micro-flocking'' state.
In contrast, the weighted distribution function for fast rotation decays at small cluster sizes $n$, but exhibits a broad peak at large $n$ [see Fig.~\ref{DCSs:fig:nPn}(b)].
This peak corresponds to the large swarms shown in Fig.~\ref{DCSs:fig:9snapshots}(c).
On increasing the particle numbers $N$, the decay of $nP\left(n\right)$ at small $n$ becomes faster, while the peak at large $n$ is shifted toward a larger $n$.
In other words, the size of the formed structure scales with the particle number $N$, indicating the emergence of macroscopic swarming patterns.
Based on Table~\ref{DCSs:table:states}, we identify this as a ``macro-flocking'' state.
The mechanism underlying the flocking behavior of dipolar active systems is quite complex already in the absence of circle swimming (that is, at $\omega_0^* = 0$, see ref.~\citen{Liao2020}): Short linear chains formed by dipolar active particles with head-tail orientation tend to align their velocities upon collisions (see Fig.~8 of ref.~\citen{Liao2020}). 
In ref.~\citen{Liao2020}, it is found that macro-flocking appears once the density increases from $\Phi = 0.23$ to $\Phi = 0.58$.
This suggests that the transition from micro- to macro-flocking may be attributed to a density-induced enhancement of particle collisions. 
For the present system of chiral dipolar active particles, a similar effect may take place: We suspect that the circle swimming of each particle leads again to an enhancement of collisions, similar to an increase of density.
This might explain why the present, chiral system exhibits macro-flocking already at smaller densities than the corresponding non-chiral system ($\omega_0^* = 0$).
We note that our observation concerning the size of the flocking pattern contrasts the behavior previously observed in systems of chiral active particles with polar interaction.
There, the size of flocking patterns is inversely proportional to the angular speed.\cite{Liebchen2017, Levis2018}
This suggests that different types of alignment interactions between active particles may significantly change the fundamental properties of flocking behavior.

\section{\label{DCSs:sec:conclusions}Conclusions}

In this work, we performed extensive Brownian dynamics (BD) simulations to investigate the pattern formation of dipolar circle swimmers dispersed on a monolayer.
To this end, we explored a wide range of active angular speeds and motilities at a fixed density below that related to MIPS in simple systems and large dipolar coupling strength.
At a sufficiently large angular speed ($\omega_0^* \gtrsim 1$) and zero motility ($v_0^* = 0$), the system undergoes a transition from a state with percolated networks into a state with Type I and Type II vortices.
Upon an increase of $v_0^*$ from zero, the Type II vortices vanish, and the system displays phase separation with the dense domain characterized by giant Type I vortices.
Based on test simulations of two anti-parallel chains composed of strongly coupled dipolar particles, we proposed a ``ring'' argument to unveil the underlying mechanism of the vortex formation and the disappearance of Type II vortices.
In contrast with our model, the vortex structures are not observed in systems of circle swimmers with ferromagnetic Heisenberg-like interactions.\cite{Liebchen2017, Levis2018}
Instead, these systems display macroscopic polar droplets, in which swimmers move coherently along a certain direction and the direction rotates in accordance with the active rotation exerting on each circle swimmer.
To further unravel the differences, we performed Brownian dynamics simulations of circle swimmers with ferromagnetic interactions decaying with a Yukawa potential (as a function of separation),\cite{Lichtner2012, Lichtner2013} and, thus, decay with the particle distance.
Our preliminary results (not shown here) do not show any vortex structures, either.
%

%% New Paragraph
%
A further increase of the motility ($v_0^* \gtrsim 80$) leads to two distinct flocking states, which can be distinguished by the cluster size distribution.
Consistent with the behavior of dipolar active Brownian particles,\cite{Liao2020} dipolar circle swimmers at zero and slow rotation display a micro-flocking state ($\omega_0^* \lesssim 3$).
In contrast, dipolar swimmers with fast rotation ($\omega_0^* \gtrsim 7$) exhibit a macro-flocking state.
This is again distinctly different from circle swimmers with short-range ferromagnetic interactions, in which the size of polar clusters decreases upon an increase of the active angular speed.\cite{Liebchen2017, Levis2018}
Hence, the type of anisotropic interactions that align particle velocities plays a vital role in determining the fundamental self-organization process for systems of circle swimmers.
For completeness, we also performed simulations for circle swimmers with truncated dipole-dipole interactions (results not shown here).
Here, the Ewald summation was not employed, and the dipole-dipole interactions were truncated at $r = 3 \sigma$.
In this case, the flocking transition is shifted to a much larger motility ($v_0^* \approx 200$).
More importantly, we did not observe any vortex patterns throughout the explored parameter regime.
This suggests that not only the angle dependency, but also the long-range character of dipole-dipole interactions is crucial for the collective behavior of dipolar active systems and, thus, should be not be neglected.
The model studied in the present paper does not account for hydrodynamic interactions between particles.
However, hydrodynamic interactions can have a profound impact on the dynamical self-assembly of active colloidal systems, such as the suppression of MIPS\cite{Matas-Navarro2014, Matas-Navarro2015, Theers2018, Schwarzendahl2019} and the emergence of global polar ordering.\cite{Alarcon2013, Delmotte2015, Yoshinaga2017, Yoshinaga2018, Hoell2018}
In systems of rotating particles, it is found that hydrodynamic interactions can induce cluster rotation, such that the overall cluster and the individual particles rotate in the same direction.\cite{Jager2013}
In the present work, we have seen that the giant Type I vortices and circle swimmers rotate in the same direction.
Hence, we expect that hydrodynamics can further promote the formation of giant Type I vortices.
Nevertheless, the detailed influence of hydrodynamics on dipolar circle swimmers remains to be unveiled by future works.
Furthermore, it is well established that mixtures of active and passive colloidal particles display fascinating collective behavior that is distinctly different from that of the corresponding one-component systems.\cite{Maloney2020, Maloney2020a,  Stenhammar2015, Wysocki2016, Wittkowski2017}
Therefore, one future direction could be to consider mixtures of circle swimmers and passive dipolar particles, whose self-assembly process can be further controlled by the proportion of species.

Finally, the majority of the research works on systems of active particles focus on their collective behavior in two dimensions (2D), whereas the real-world suspensions of self-propelled colloids are often in three dimensions (3D).
The dimensionality may play an important role in the motility-induced phenomena, such as the critical motility for MIPS\cite{Stenhammar2014} and the critical exponents for flocking transition.\cite{Ginelli2016}
In the present 2D model of dipolar circle swimmers, we assume that the particles are confined to a flat surface, and the rotation axis for each particle is restricted to be normal to the surface.
It will be interesting to discover the similarities and differences in our model system when moving from 2D to 3D.

\section*{Conflicts of interest}

There are no conflicts to declare.

\appendix
\begin{figure}[!thbp]
  \centering
  \includegraphics[width=0.8\linewidth]{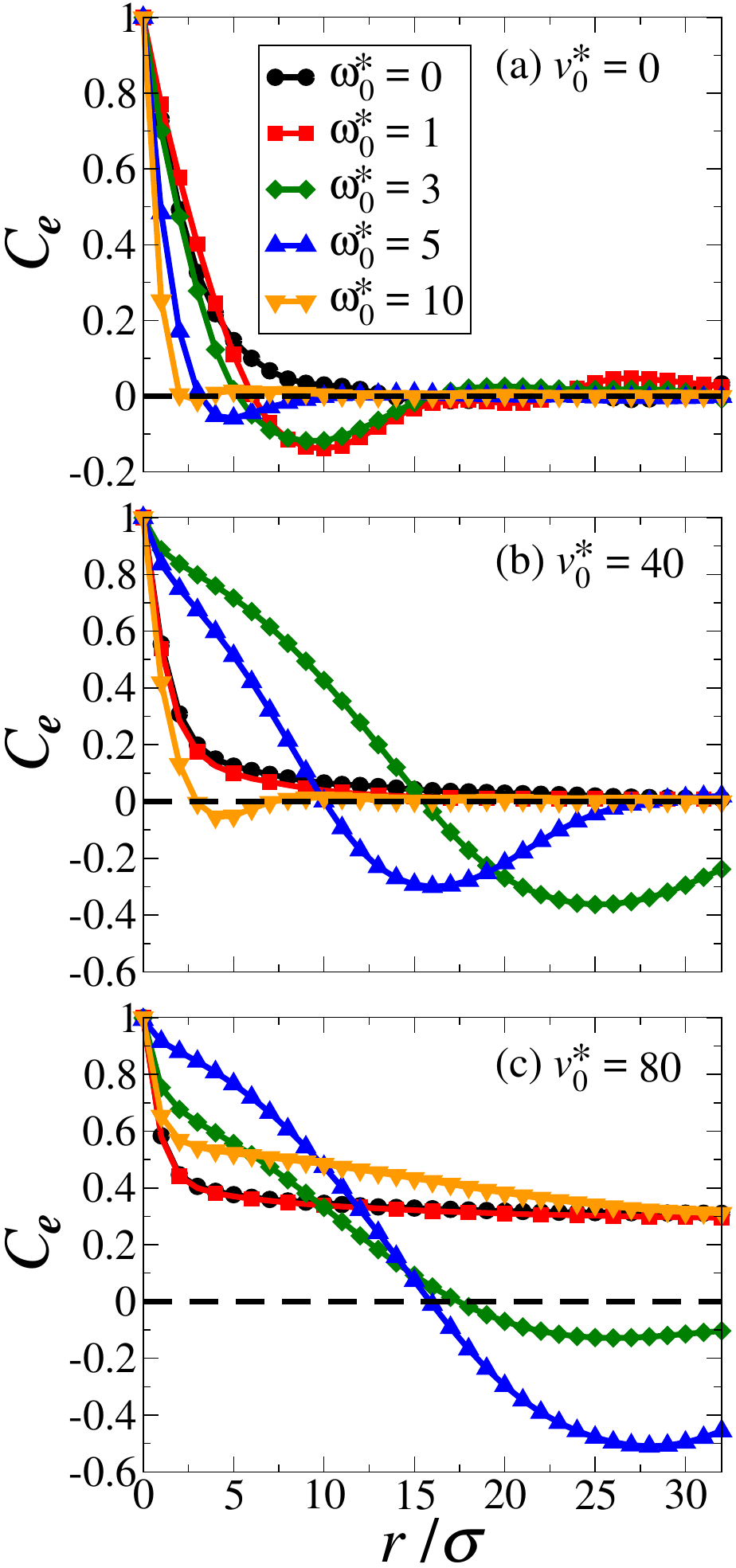}
  \caption{Orientational correlation function of distance $C_e\left(r\right)$ for the motility $v_0^* = 0$ (a), $40$ (b), and $80$ (c) and the angular speed $\omega_0^* = 0$ (black dots), $1$ (red squares), $3$ (green diamonds), $5$ (blue triangles up), and $10$ (orange triangles down).}
  \label{DCSs:fig:Ce}
\end{figure}
\section{\label{DCSs:sec:Ce_of_r}Orientational correlation function}
To characterize the vortex structure, we calculate the orientational correlation function $C_{\boldsymbol{e}}(r)$ defined in Eq.~\eqref{DCSs:eqn:Ce}.
Results are shown in Fig.~\ref{DCSs:fig:Ce}.
For passive dipolar particles ($v_0^* = 0$ and $\omega_0^* = 0$), $C_{\boldsymbol{e}}(r)$ decays with $r$ and reaches zero at $r / \sigma \approx 10$ [see Fig.~\ref{DCSs:fig:Ce}(a)].
Upon an increase of $\omega_0^*$ from zero to $1$, $C_{\boldsymbol{e}}(r)$ decays faster than the case where $\omega_0^* = 0$ and displays a negative correlation with a minimum $C_{\boldsymbol{e}}(r) \approx -0.15$ at $r / \sigma \approx 10$.
As have been discussed in Sec.~\ref{DCSs:sec:vortex}, the negative correlation indicates the emergence of vortex structures.
Further, the distance corresponding to the minimum of $C_{\boldsymbol{e}}(r)$ determines the average vortex diameter [see Eq.~\eqref{DCSs:eqn:de}], and the minimum value of the correlation function $C_{\boldsymbol{e}}(d_{\boldsymbol{e}}) \approx -0.15$ represents the prominence of the vortex structures.
As the angular speed increases from $1$ to $10$, both the vortex diameter $d_{\boldsymbol{e}}$ and the significance of the vortex structures decrease, indicating that the vortex structures occur at a sufficiently large angular speed $\omega_0^* \approx 1$ and are inhibited by further increasing $\omega_0^*$.
Comparing the results for $v_0^* = 0$ and $v_0^* = 40$, the vortex structures emerge at $\omega_0^* = 1$ for $v_0^* = 0$, whereas for $v_0^* = 40$ the vortices appears at a larger angular speed $\omega_0^* = 3$ [see Fig.~\ref{DCSs:fig:Ce}(b)].
In particular, the vortex size $d_{\boldsymbol{e}}$ becomes much larger and is close to the half of the side length of the simulation box ($L / 2 \approx 33.69 \sigma$), suggesting that there might be finite-size effect for $d_{\boldsymbol{e}}$.
In other words, $d_{\boldsymbol{e}}$ may scale with the total number of particles $N$, which requires further investigation.
Nevertheless, since the long-range character of dipole-dipole interactions requires expensive computational resources, the present simulations are limited to the order of $10^3$ particles.
A further increase of $\omega_0^*$ from $3$ to $10$ causes the giant vortices to break into smaller vortices.
At high motilities such as $v_0^* = 80$, $C_{\boldsymbol{e}}(r)$ drastically decays at short distances $r / \sigma \leq 3$ and gradually reaches a value close to $0.3$ at $r = L / 2$ for zero, small and large angular speeds ($\omega_0^* = 0, 1$ and $10$).
The non-vanishing, positive correlation function indicates the emergence of global orientational order, which is discussed in detail in Sec.~\ref{DCSs:sec:result3}.
At the intermediate angular speeds ($\omega = 3$ and $5$), the minimum of $C_{\boldsymbol{e}}(r)$ appear at $r \lesssim L/2$.

\section*{\label{DCSs:sec:acknowledgement}Acknowledgements}

This work was financially supported by the Deutsche Forschungsgemeinschaft under GRK 1524 (DFG No. 599982).

%\section*{Notes and references}

%%%END OF MAIN TEXT%%%

%The \balance command can be used to balance the columns on the final page if desired. It should be placed anywhere within the first column of the last page.

%\balance

%If notes are included in your references you can change the title from 'References' to 'Notes and references' using the following command:
%\renewcommand\refname{Notes and references}

%%%REFERENCES%%%
\bibliography{DCSs} %You need to replace "rsc" on this line with the name of your .bib file
\bibliographystyle{rsc} %the RSC's .bst file

\end{document}